# Observational Evidence Revises Presumed Large Ozone Worsening from Nitrogen Oxides Cuts

Xiang Weng[1,2,3], Xiao Lu[1,2]*, Jiawei Li[4], Grant Forster[3], Jessica Chapman[3,5], Beckie George[3,5], Yunbo Lu[1,2], Guowen He[1,2], Haofan Wang[6], Jingcheng Lai[1,2], Peer Nowack[7,8]

[1]School of Atmospheric Sciences, Sun Yat-sen University, and Southern Marine Science and Engineering Guangdong Laboratory (Zhuhai), Zhuhai, 519082, China;
[2]Guangdong Provincial Observation and Research Station for Climate Environment and Air Quality Change in the Pearl River Estuary, Zhuhai, 519082, China;
[3]School of Environmental Sciences, University of East Anglia, Norwich, NR4 7TJ, U.K;
[4]State Key Laboratory of Earth System Numerical Modeling and Application, Institute of Atmospheric Physics, Chinese Academy of Sciences, Beijing, 100029, China;
[5]Tyndall Centre for Climate Change Research, University of East Anglia, Norwich, NR4 7TJ, UK;
[6]College of Resources and Environment, Chengdu University of Information Technology, Chengdu, 610225, China
[7]Institute of Theoretical Informatics, Karlsruhe Institute of Technology, Karlsruhe, 76131, Germany;
[8]Institute of Meteorology and Climate Research (IMKASF), Karlsruhe Institute of Technology, Karlsruhe, 76131, Germany

*Correspondence to: Xiao Lu (luxiao25@mail.sysu.edu.cn)

## Abstract

Many air quality models indicate that rapid reductions in nitrogen oxides ($NO_x$), without comparable controls on volatile organic compounds, have worsened summertime ozone pollution in urban China, producing a short-term strong ozone penalty. Other models, however, simulate the opposite response, suggesting that cutting down $NO_x$ has already helped mitigate ozone pollution. This contradiction obscures understanding of atmospheric chemistry and weakens guidance on control policy design. Here, we reconcile this disagreement and reveal the underestimated benefits of $NO_x$ emission reductions using a machine learning framework integrated with an observational constraint. We first constrain ozone responses under a 30% $NO_x$ reduction, comparable to the magnitude of $NO_x$ emission declines across major Chinese city clusters between 2015 and 2023. The constrained results indicate that ozone decreases prevail across urban China, with only small increases mainly in July 2015. This challenges the widespread ozone worsening that many models predict. We then extend the constraint across 10-60% $NO_x$ reductions, establishing its use for rapid ozone sensitivity diagnosis without exhaustive scenario modeling. This diagnosis shows that sustained $NO_x$ control increasingly favored ozone mitigation during 2015-2023, benefiting a growing share of China's population. These results underscore that continued $NO_x$ reductions can deliver larger ozone mitigation benefits than many models suggest.

## Introduction

Air quality models are crucial for understanding atmospheric processes and composition[1–5]. Yet, like all models, they are inevitably subject to uncertainty. In fact, commonly used numerical models even disagree in how anthropogenic emission changes may have contributed to surface ozone pollution in urban areas. This is particularly consequential for China, where summertime ozone has worsened markedly over the past decade[6–8], especially across densely populated urban regions[9,10]. Many model-based studies have argued that sharp reductions in emissions of nitrogen oxides ($NO_x$) without comparable regulations on volatile organic compounds (VOCs) were an important driver of the rising summertime ozone in urban China, especially before 2017[11–15] when $NO_x$ emissions were high relative to levels in recent years[16]. This is a phenomenon known as the $NO_x$-related ozone penalty, which can occur in urban regions that are saturated with $NO_x$ owing to large emissions[17,18]. Under such $NO_x$-saturated conditions, VOC-prioritized or combined VOC-$NO_x$ controls are often recommended for near-term ozone mitigation[1,19], whereas sustained $NO_x$ reductions are needed to shift the atmosphere to conditions in which further $NO_x$ controls reduce rather than increase ozone[20].

This interpretation, however, is not uniformly supported across models. Studies using different models[21,22] have simulated summertime ozone decreases across many urban areas in China under $NO_x$ reductions alone, even for earlier years including 2017 and 2018. Evidently, differences in simulated meteorology and input emissions can contribute to modeling uncertainty[23–25], but the large disagreement in simulating ozone responses to $NO_x$ changes is likely driven predominantly by how models represent ozone chemistry differently, i.e., their embedded chemical mechanisms. This is supported by model intercomparison studies using different chemical mechanisms over North America[26], Europe[27], and more recently Korea[28] and urban China, where two widely used chemical mechanisms in WRF-Chem, a regional air quality model, simulate divergent ozone responses to $NO_x$ reductions[29,30]. These paradoxical results create a practical dilemma for policy implementation. If $NO_x$ reductions worsen ozone, deeper cuts in $NO_x$ complemented by effective VOC regulation should be pursued; if $NO_x$ reductions, however, have already translated into ozone mitigation, then continuing existing measures should suffice. This contradiction exemplifies a broader bottleneck in air pollution modeling: how to provide robust guidance for policymakers to protect public health when models disagree. Scientifically, it obscures our understanding of ozone chemistry and of how atmospheric composition evolves in response to human-induced emission changes. As a major air pollutant associated with an estimated 1.407 million deaths worldwide per year from long-term exposure[31] and with crop-yield losses that threaten food security[32], as well as an important greenhouse gas[33], uncertainty in ozone modeling exerts far-reaching impacts. Therefore, constraining this uncertainty is of paramount importance.

Efforts to reduce modeling uncertainty have conventionally focused on improving chemical mechanisms, emissions, and physical parameterizations, guided by new observations and scientific understanding[34]. This progress is essential, but it is gradual and can still leave large structural uncertainty within air quality models. Statistical[35] and more recently machine learning based bias correction[36–40], offers a complementary route. Here, a machine learning algorithm learns and corrects the historical mismatch between simulated and observed ozone, then applying this learned correction to other scenarios such as future or counterfactual emission changes. The assumption is that a correction learned under present-day conditions remains valid when emissions change. This assumption, however, may fail[41], particularly under large $NO_x$ reductions, where ozone sensitivity to $NO_x$—and therefore the model bias itself—can shift. It resembles an extrapolation challenge in machine learning[42], where an algorithm struggles to predict conditions absent from the training data. As a result, statistical or machine learning based bias correction may produce an ostensibly corrected ozone concentration baseline, often demonstrated by the exceptionally close agreement with observations without delivering any concrete improvement in predictions for emission reduction scenarios. It may even overcorrect the model whilst suppressing its inherent physically meaningful signals[43], thereby producing unreasonable ozone responses to $NO_x$ reductions. This limitation is difficult to test directly, because counterfactual or future emission scenarios cannot be observed. Furthermore, bias-correction frameworks for ozone are often developed for a single model at a time, leaving their transferability across models unexamined.

To address these gaps, we develop an efficient machine learning framework tailored to narrow the discrepancies in predicted ozone responses to $NO_x$ emission changes among numerical models, based on the concept of observational constraints applied in climate science[44,45]. This framework reveals the underestimated benefits of $NO_x$ reductions for curbing the severity of ozone pollution over urban China. It subsequently enables a data-driven diagnosis of ozone responses across a wide range of $NO_x$ emission reductions and quantifies the effect of long-term historical $NO_x$ changes on ozone pollution in China at an exceptionally low computational cost, an approach that can easily be deployed at a global scale.

## Emulation of divergent ozone sensitivity predictions by different models

The framework for constraining uncertainties in simulated ozone responses to $NO_x$ emission reductions consists of two main steps: emulation and reconciliation. In the emulation step, we use Gaussian process regression (GPR) as the learning function together with the spatial-expansion strategy described in Methods, to iteratively reproduce the ozone responses to $NO_x$ reductions simulated by five WRF-Chem chemical mechanisms, with each mechanism treated as an individual numerical model in this study. This step is crucial as it tests whether a data-driven function can learn each model's intrinsic and complex—albeit uncertain—behavior for ozone sensitivity predictions. For training the function, both predictand ($y$, i.e., ozone) and predictor are collected from the model outputs under the retrospective baseline scenario, in which the models reconstruct historical ozone during July 2015, 2017 and 2019 using corresponding meteorological and emission inputs (see Methods for details). Training is restricted to the baseline simulations because the same framework is later retrained with observations to derive the observational relationship (i.e., building the observational constraint) for reconciling the divergent numerical model predictions. This design, therefore, requires the learned relationship to extrapolate ozone responses under unseen $NO_x$ reduction scenarios, which is extensively evaluated in the beginning of this study (see Fig. 1).

In the reconciliation step, as aforementioned, both the simulated predictand and predictor are replaced with observations to retrain the GPR. The resulting observational relationship is then used to constrain the model predictions based on their simulated changes in the predictor under $NO_x$ emission reductions. Therefore, this two-step approach rests on the premise that much of the model disagreement in ozone responses arises from differences in the represented ozone chemistry, which are learned during the previous emulation step and then constrained by the observational relationship in this step. In brief, we evaluate this premise using a predictor-swap test, in which the predictor changes under $NO_x$ reductions are exchanged between the two emulation functions that learn the divergent ozone responses. Further technical details for this test and the overall framework are provided in Methods, with a graphical illustration of the framework shown in Supplementary Fig. 1.

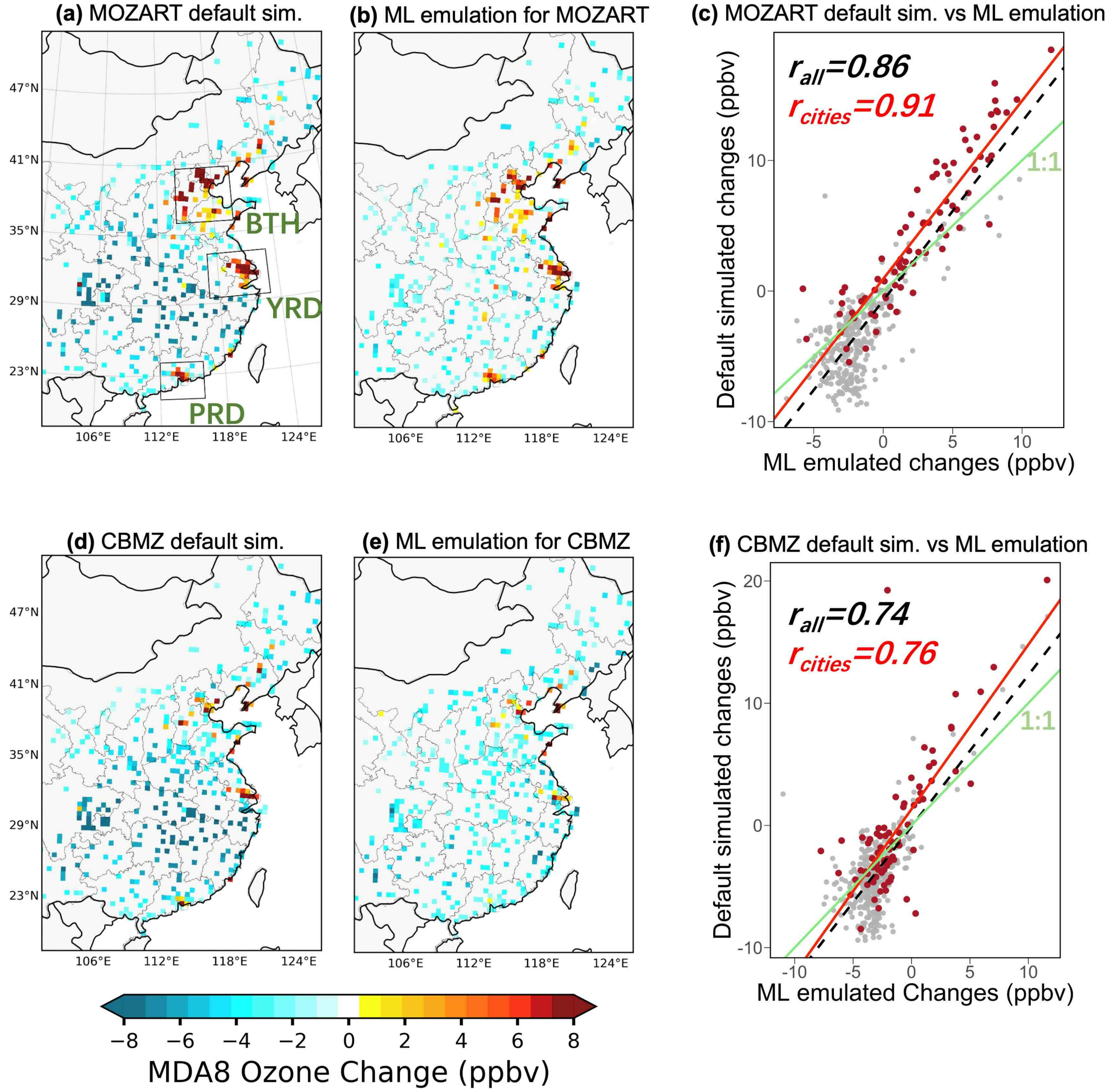

**Fig. 1 | Machine learning emulation of divergent ozone responses to 30% $NO_x$ emission reductions simulated by MOZART and CBMZ).** a–c, MOZART results: **a**, Default simulated ozone response to 30% $NO_x$ emission reductions in July 2019, with the domains of the three city clusters labeled (Beijing-Tianjin-Hebei, BTH; Yangtze River Delta, YRD and Pearl River Delta, PRD). **b**, Corresponding emulation **c**, Spatial correlation between default simulation (a) and emulation (b). In this scatter plot, gray points indicate grid cells outside the three city clusters, and dark-red points indicate grid cells within the clusters. Pearson correlation coefficients for all grid cells across China (including both gray and red points) and for the city clusters are inserted in the top-left corner with black and red, respectively. Black dashed lines show linear fits for all grid cells, red solid lines show fits for city-cluster grid cells only, and green lines indicate the 1:1 relationship. **d-f**, As in a-c, but for CBMZ.

Fig. 1 demonstrates the skill of the machine learning function in emulating the contradictory ozone responses to 30% $NO_x$ emission reductions as simulated by the two commonly used chemical mechanisms in WRF-Chem, MOZART and CBMZ. MOZART predicts that, despite $NO_x$ emission reductions, ozone levels increase considerably by 5-8 ppbv over the city clusters of China (Fig. 1a). Notably, this pattern is well replicated by the GPR (Fig. 1b), with Pearson correlation coefficients of 0.86 over China and 0.91 within the three clusters (Fig. 1c) where the modeling discrepancies are most notable. With the same setup, GPR also reproduces the CBMZ responses (Fig. 1e-f), predicting weaker ozone increases and even decreases in certain locations over the same city clusters.

These results bear great significance. Firstly, GPR is trained only on model output from baseline simulations (July 2015, 2017 and 2019), yet it successfully reproduces the two contrasting predicted ozone changes driven by a large unseen counterfactual 30% $NO_x$ reduction in July 2019 (Fig. 1b and e). This highlights the extrapolation skill of the learning architecture, providing confidence in its application to unseen $NO_x$ reduction scenarios. Remarkably, only a single predictor, simulated surface $NO_2$, is used to achieve the extrapolation in the context of emulating the complex nonlinearity of ozone chemistry. This ability is largely enabled by the spatial-expansion design (Methods), which trains each target grid cell with neighbouring samples that cover a broader range of ozone-$NO_2$ conditions, including situations resembling those reached under $NO_x$ reductions. GPR then fits a smooth relationship across these samples, supporting sufficient extrapolation to this unseen 30% $NO_x$ reduction scenario (Supplementary Fig. 2). The approach of using-$NO_2$-alone also concurs with a recent study suggesting that $NO_2$ is a reliable predictor of ozone sensitivity[46], as well as directly indicative of changes in $NO_x$ emissions[47]. It also opens a broader research avenue. As $NO_2$ is widely measured, the framework can be readily applied to emulate and then observationally constrain model responses in other regions worldwide. Furthermore, other numerical models can also be emulated using the same machine learning architecture (see Extended Data Fig. 1), including the intentionally distorted model termed "CB-MOD" (MOD means modified), or colloquially named the "wrong" model herein. It is a modified CBMZ mechanism in which default VOC reactivities were effectively suppressed to substantiate the underlying causes of the discrepancies between CBMZ and MOZART from our previous study[30] (Methods). This broader performance emphasizes the versatility of the framework across diverse models.

## The underestimated benefit of $NO_x$ reductions in easing ozone pollution

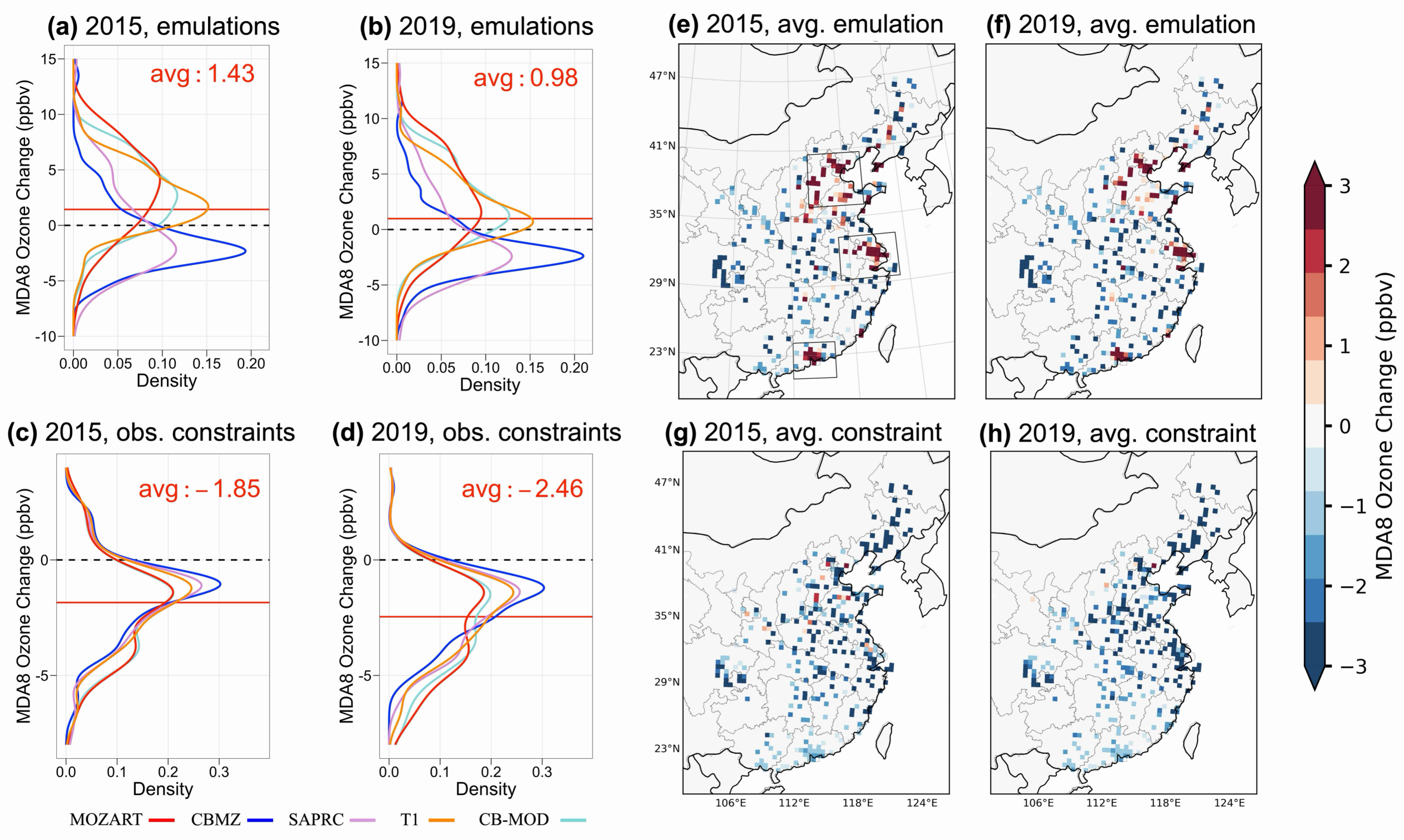

**Fig. 2 | Emulated and observationally constrained ozone responses to 30% $NO_x$ emission reductions. a**,**b**, Density distributions of the machine-learning-emulated ozone responses from the five chemical mechanisms within the three city clusters in July 2015 and July 2019, respectively. **c**,**d**, Corresponding observationally constrained predictions. Red horizontal lines indicate mean ozone changes, with values shown in red at the top right corner. **e** and **f**, Spatial distributions of the averaged emulations for ozone responses of the five mechanisms in July of 2015 and 2019, respectively. **g** and **h**, the same for the observation constrained results. Because the observational constraint is trained on observations that are subject to greater unaccounted influences and therefore require sufficient adjacent grids for extrapolation (Methods), predictions are not made on the grids without enough adjacent information . For consistency, those same grids from e and f are discarded. No grids within city clusters are excluded.

We next derive the observational ozone-$NO_2$ relationship (termed the observational constraint hereafter) to constrain the uncertainties in predicted ozone responses to 30% $NO_x$ reductions by models for July of 2015 and 2019, respectively. As in 2019, GPR can also emulate 2015 ozone responses across all chemical mechanisms (Extended Data Fig. 2). Five models in total are reconciled, as shown by the convergence of predictions (Fig. 2c, d) from the initially divergent emulated responses over the city clusters (Fig. 2 a, b). This highlights the framework's robust reconciliation capability and supports its central premise that much of the model spread arises from differences in the representation of ozone chemistry (the responsive relationship between $O_3$ and $NO_2$ as emulated by the learning function), rather than from any inconsistency in simulated $NO_2$ (predictor) changes in response to the $NO_x$ emission perturbations (Methods).

In contrast to the averaged ozone increases emulated by all the machine learning emulators, the constrained results show averaged ozone decreases in July of both 2015 (-1.85 ppbv; Fig. 2c) and 2019 (-2.46 ppbv; Fig. 2d) over the city clusters with 30% $NO_x$ reductions. Spatially, the default emulations produce far more pronounced and extensive averaged ozone increases stretching over the city clusters (Fig. 2 e, f), while the constraints indicate widespread ozone decreases, with only localized increases remaining mainly over BTH and YRD (Fig. 2g, h). These observational constraints underline the important message that effectively reducing $NO_x$ emissions by 30% can considerably ease ozone pollution. Even in 2015 when $NO_x$ emissions were relatively more intense, observational constraints indicate that only sporadic areas within BTH and YRD would face worsening ozone with this level of $NO_x$ reduction. This is despite the higher intensity of $NO_x$ emissions in 2015 possibly pushing the ozone sensitivity toward a more $NO_x$-saturated regime wherein widespread elevated ozone would theoretically be expected with $NO_x$ reductions alone, especially over these high-$NO_x$-emission city clusters. The constrained results particularly challenge the ozone chemistry interpretations of some of the most widely used models such as MOZART, which simulates the largest $NO_x$-reduction-induced ozone penalties with ozone increases exceeding 8 ppbv in some areas. More importantly, results here provide a more nuanced view of the common model-based interpretation that substantial policy-driven reductions in $NO_x$ emissions without equivalent regulations on VOC emissions, have inadvertently worsened urban ozone in China[10–12,14,15]. Although ozone still increased in parts of BTH and YRD given a 30% $NO_x$ reduction in 2015, the observational constraint points toward a higher efficacy of the $NO_x$ reduction strategy in limiting ozone deterioration than implied by many of these models, as their mean emulated results show stronger and more widespread ozone increases.

The stark contrast between the default emulation average and the constrained result, exemplifies that "model democracy"[48] may not always lead to reliable predictions. That is, the unweighted average of all participating models (Fig. 2e, f) does not guarantee robust predictions given that their averaged increases are now refuted by the more closely aligned constrained predictions of limited ozone penalties.

The capability of the observational constraint to even correct the deliberately distorted "wrong" model, i.e., CB-MOD (Fig. 2c, d), is noteworthy. This showcases that a robust observational constraint can effectively reduce structural model errors, even when those errors are large (intentionally inflated here). It is also reasonable to deduce that, even with extremely "wrong" models included, such as CB-MOD here, the prediction by the observational constraint may not be swayed. Instead, it may still anchor the prediction toward ozone reductions.

We also address the potential influence of meteorology on the observed ozone-$NO_2$ relationship as well as the likely discrepancies in fitted relationships derived from different configurations in GPR (see Method). Reassuringly, constraints considering the meteorological effects (Supplementary Fig. 3 and Supplementary Fig. 4) and different configurations (Supplementary Fig. 5 and Supplementary Fig. 6) produce results consistent with those presented here (Fig. 2c, d, g, h), demonstrating the robustness of our default machine learning fits.

## An efficient diagnosis of ozone sensitivity to the full range of nitrogen reductions

The observational constraint underscores that a 30% reduction in $NO_x$ emissions would not trigger widespread and strong ozone increases across the city clusters in either 2015 or 2019. A key follow-up question is how ozone may respond to incremental or larger $NO_x$ reductions. This is particularly relevant if a 30% cut is deemed to be overly ambitious or lax in some regions or future emission control pathways. We therefore develop a rapid ozone sensitivity diagnosis based on the observational constraint to estimate ozone responses across $NO_x$-reduction levels from 10% to 60% in fine increments (e.g., 5% here), without running a full suite of computationally expensive yet uncertain numerical simulations. This is supported by the remarkable ability of the GPR emulator to replicate divergent responses of MOZART and CBMZ, even under a 60% $NO_x$ reduction (see Supplementary Fig. 8). By diagnosing ozone responses across the full reduction range and on a grid-by-grid basis (Methods), this approach adaptively and efficiently derives ozone sensitivity across the study domain at a fine incremental resolution. To reiterate, because the constraint relies only on surface $NO_2$—a widely measured chemical species—this diagnosis offers a transferable route for efficient and accurate tracking of long-term historical and future ozone sensitivity around the world.

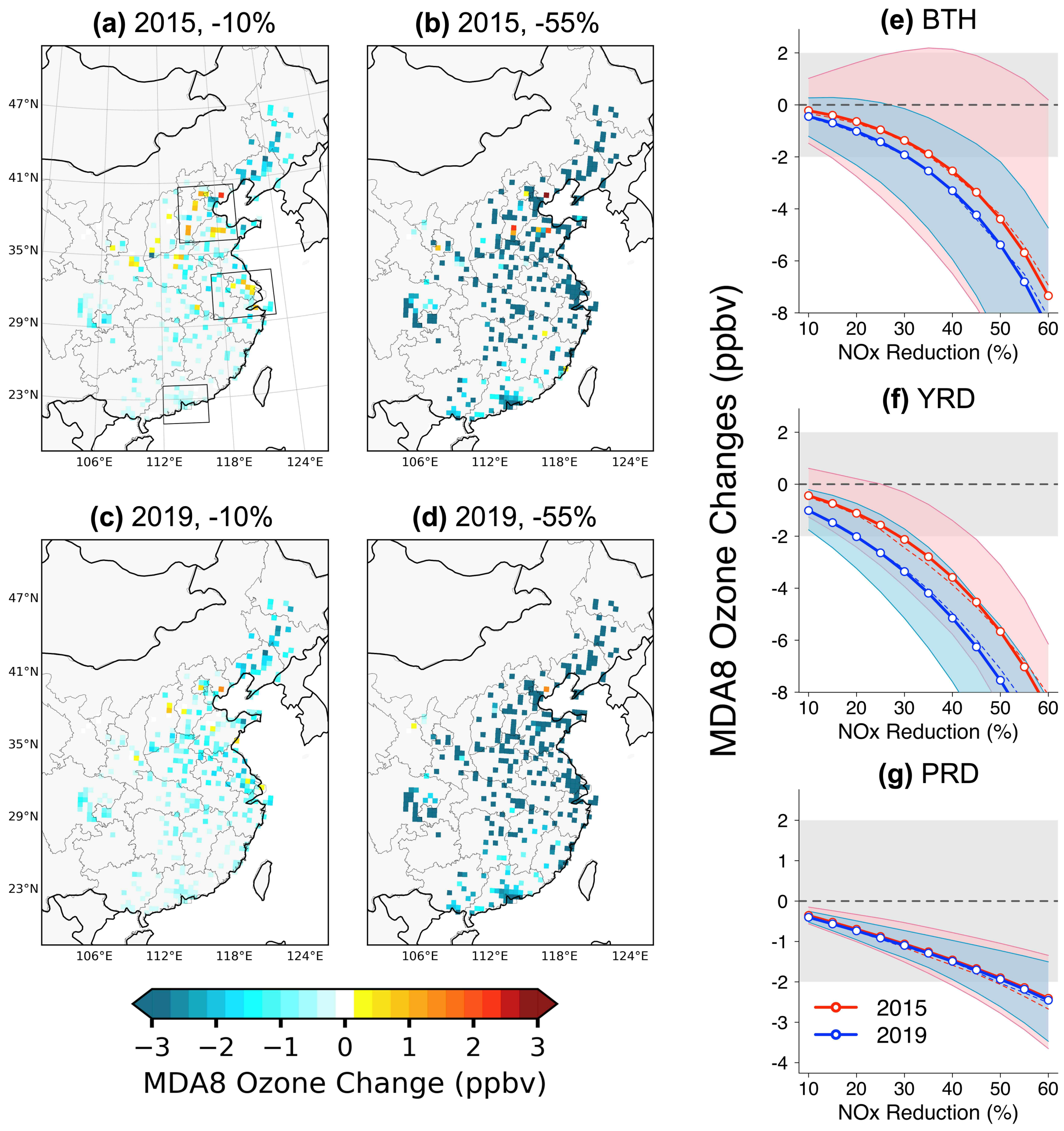

**Fig. 3 | Predicted ozone responses to prescribed $NO_x$ emission reductions. a**, **b**, Spatial distributions of monthly mean MDA8 ozone changes in July 2015 in response to $NO_x$ emission reductions by 10% and 55%, respectively. **c**, **d**, As in a,b, but for July 2019. Spatial distributions for remaining perturbations are available in Extended Data Fig. 3. **e-g**, Ozone changes responding to $NO_x$ emission reductions from 10% to 60% for BTH, YRD and PRD, respectively. Solid lines represent the regional mean ozone responses under a given $NO_x$ reduction, dashed lines (mostly overlapping with the solid lines) show the median and shading around these lines indicates the $10^{th}$-$90^{th}$ percentile range. Red lines and pink shading denote predictions for July 2015, and blue lines and light blue shading denote those of July 2019. The shaded gray bands indicate the range of ozone changes from -2 to 2 ppbv, shown to facilitate cross-region comparison.

The rapid diagnosis shows that at the low end, a 10% $NO_x$ reduction still produces small ozone increases of 0.5 to 2 ppbv over areas inside BTH and YRD with a broader and stronger response in 2015 than in 2019 (Fig. 3a and c), and the increases are far less dramatic than the default MOZART simulations suggest (Supplementary Fig. 11, Supplementary Fig. 12). In contrast to 2019, even under a 55% reduction, parts of BTH are still estimated to experience slightly worsened ozone in 2015. These results reflect stronger $NO_x$-saturated conditions over BTH and YRD in earlier years, broadly consistent with satellite-based studies that inferred ozone sensitivity.[49,50]

The contrast among the three city clusters is noticeable. BTH stands out as the most $NO_x$-saturated region, with small ozone increases persisting across a wide range of $NO_x$ reductions (Fig. 3e, Extended Data Fig. 3), probably owing to its relatively high $NO_x$ emissions[16]. By contrast, such persistent ozone increases are absent over YRD, where ozone increases tip almost fully to decreases once $NO_x$ is cut beyond 30% (Fig. 3f, Extended Data Fig. 3). Interestingly, the PRD in southern China shows the clearest directional response, with ozone decreasing under $NO_x$ reductions in both years (Fig. 3g). Yet, these decreases are modest compared with those in BTH and YRD under stronger controls.

Taken together, the diagnosis here depicts a regionally heterogeneous but broadly favorable response to $NO_x$ controls. Small reductions in $NO_x$ emissions might inadvertently yet only slightly elevate ozone by 0.5 to 2 ppbv over BTH and YRD. But $NO_x$ reductions do not appear to be a prevailing cause for the drastic ozone worsening, as explored further below. Instead, cutting $NO_x$ generally favors ozone mitigation, with increasing efficacy from 2015 to 2019 across all three city clusters. This transition is consistent with observation-based studies using both satellite retrievals and ground-based in situ measurements[49–51].

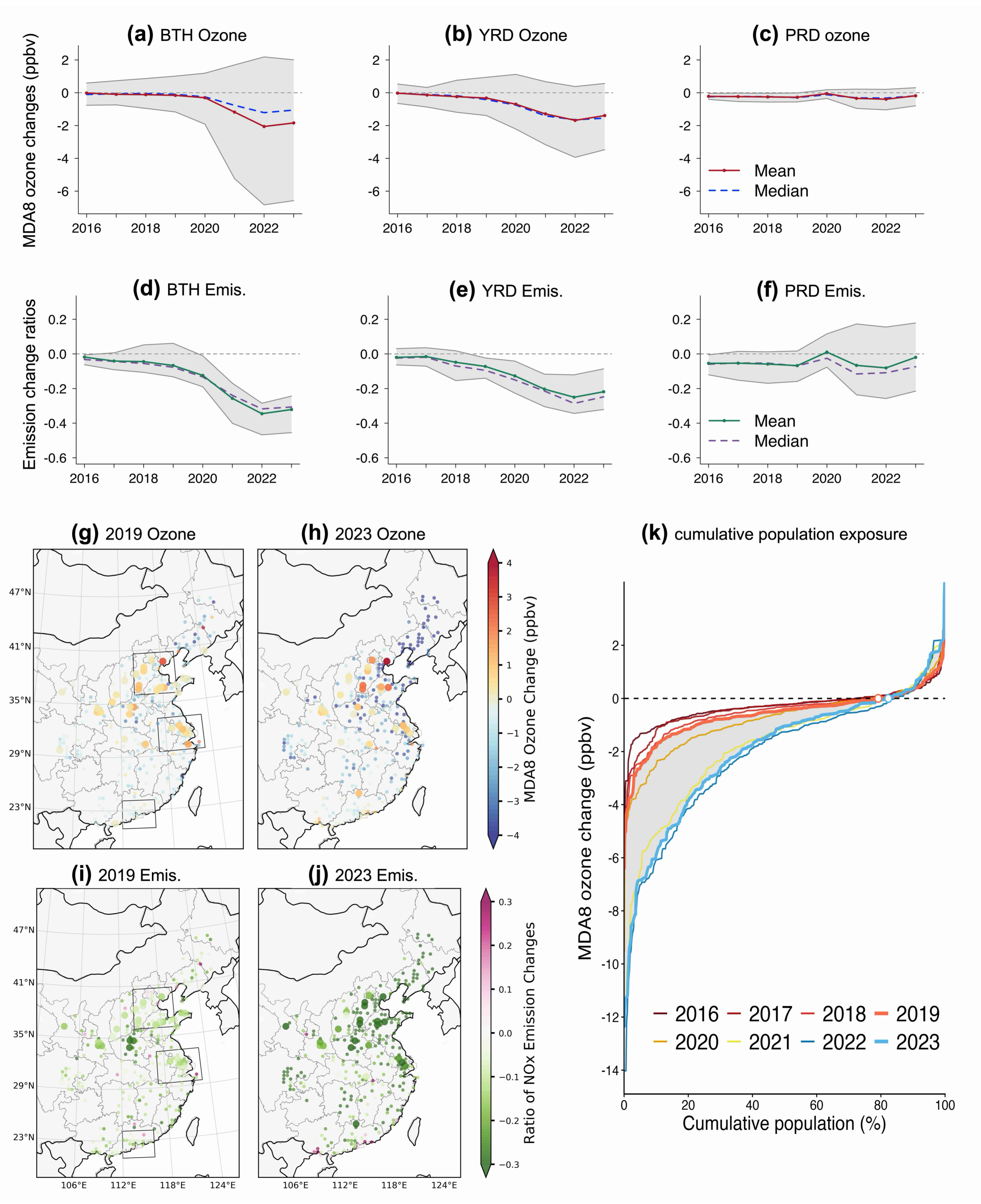

**Fig. 4 | Observational-constraint-predicted ozone responses to $NO_x$ emission changes from 2016 to 2023 relative to 2015. a-c**, Regional monthly MDA8 ozone changes in July 2016-2023 relative to July 2015 for BTH, YRD and PRD, respectively. Red solid lines show the regional mean, blue dashed lines show the median, and gray shading denotes the $10^{th}$-$90^{th}$ percentile range. **d-f**, Regional $NO_x$ emission changes relative to July 2015 for BTH, YRD and PRD, respectively. Green solid

lines are regional means, purple lines are medians, and gray shading represents the 10$^{th}$-90$^{th}$ percentile range. **g**, **h**, Spatial distributions of predicted monthly mean MDA8 ozone changes in July 2015 when $NO_x$ emissions are changed to their July 2019 and July 2023 levels, respectively. Each point marks the center of a WRF-Chem grid cell. Larger points indicate grid cells where ozone increases despite reductions in $NO_x$ emissions. **i**,**j**, As in g,h, but for ratio of $NO_x$ emission changes relative to July 2015. Larger points correspond to the same grid cells highlighted in g,h. Spatial distributions for all years (2016-2023) are available in Extended Data Fig. 4 (ozone) and Extended Data Fig. 5 (Emission).**k**, Cumulative population distributions of changes in MDA8 ozone relative to July 2015 given $NO_x$ emission changes, with curves color-coded by year. Circular markers on the zero horizontal line denote the population thresholds separating ozone increases from decreases in 2019 and 2023. Gray shading indicates the gap between the 2019 and 2023 curves. Only the population within the analyzed grid cells shown in g–j was included in this population-exposure analysis.

Without running numerical models, we then apply the rapid diagnosis approach to assess how $NO_x$ emission changes alone may have shaped China's ozone evolution from 2016 to 2023 with the ozone baseline set July 2015. The results here reiterate the importance of sustained $NO_x$ emission controls. Larger $NO_x$ reductions in 2023 than in 2019 (Fig. 4i and j) lead to more widespread ozone decreases across the city clusters (Fig. 4g and h), supporting $NO_x$ reduction as an effective long-term strategy. Localized ozone increases nevertheless remain, particularly over BTH, but their magnitude is small, reaching at most about 2-3 ppbv (Fig. 4a, g and h). This residual ozone penalty is in accordance with the sensitivity diagnosis in Fig. 3, given that average $NO_x$ reductions over BTH in 2023 are still only around 30% relative to 2015 levels (Fig. 4d), a range in which ozone increases can persist in parts of this strongly $NO_x$-saturated region (Fig. 3e).

From a policy-relevant perspective, continued reductions—but not as large as some models would have recommended—are needed to sustain more widespread and effective ozone mitigation. To further illustrate this point, the slight rebound of $NO_x$ emissions in 2023 relative to 2022 across the three city clusters (Fig. 4d-f) is estimated to re-exacerbate ozone pollution in PRD, where ozone is least sensitive to $NO_x$ changes among the three city clusters. Although these increases are small, they nonetheless emphasize the need for consistent $NO_x$ emission reductions.

We further assessed how $NO_x$ emission changes alone affected population exposure to ozone across these years, recognizing that ozone-related health burdens depend on the population exposed, not only on gridded concentration changes. Overall, $NO_x$ controls deliver a clear and growing population-level benefit: an increasing fraction of the population experienced ozone reductions, as shown by the shift of the cumulative exposure distributions toward negative ozone changes between 2019 and 2023. Because ozone exposure carries a substantial mortality burden, this shift underscores a health-relevant benefit of emission controls that models predicting a widespread ozone penalty would have missed. Although the inferred ozone penalty from $NO_x$ reductions is substantially weaker than those models suggest, the smaller population still exposed to ozone increases experienced slightly larger increases in 2023 than in 2019. This is evident from the 2023 curve lying to the left of the 2019 curve in the positive-ozone-change range (Fig. 4k), consistent with the stronger increases over sporadic hotspots in the spatial maps (Fig. 4g, h). This is likely that these hotspots remain more $NO_x$-saturated than those adjacent areas where ozone mitigation occurs. Furthermore, because ozone concentrations were already high in the July 2015 baseline[9], worsening in these hotspots means greater health risks for the population there. These results therefore reinforce the importance of sustaining $NO_x$ reductions, while strengthening controls in these hotspots and coordinating them with broader precursor regulations to further protect human health.

For future research, our lightweight framework opens opportunities to efficiently assess ozone responses under emerging air quality challenges, including wildfire-driven ozone episodes[52], increasing contributions of natural sources such as soil $NO_x$ and climate-related ozone penalties, for which a recent study suggests that $NO_x$ reduction could become more effective under a warming climate[53]. Finally, where available, additional measurements such as formaldehyde (HCHO) or speciated VOCs would provide valuable constraints on VOC-ozone chemistry and allow the framework to examine ozone responses to VOC reductions.

## Methods

### Framework for constraining uncertainties in simulated ozone sensitivity

Here, we provide detailed explanations of the machine learning framework used to constrain uncertainty in ozone sensitivity to $NO_x$ emission changes, as simulated by numerical air quality models. As summarized in the main text, the framework consists of two steps: emulation and reconciliation. A graphical summary of these steps is shown in Supplementary Fig. 1.

For emulation, we combine training of a Gaussian Process Regression (GPR; detailed in a later section) with a spatial-expansion modeling strategy (described in the following section) to derive a model-specific emulation function, here generically termed $f_m$. In this step, numerical-model-simulated ozone is used as the predictand ($y_M$), and simulated surface $NO_2$ as the predictor ($x_M$). Both of these simulated variables are from baseline scenario. In these mathematical expressions, a lowercase subscript $m$ denotes quantities related to machine learning, whereas an uppercase subscript $M$ denotes quantities originating from numerical models. The aim is for the emulator to replicate the ozone responses to $NO_x$ reductions simulated by each WRF-Chem chemical mechanism, with each mechanism treated as an individual numerical model in this study (denoted as $M_1$, $M_2$, …). As the main text outlined, this step validates whether the purely data-driven function, $f_m$ ($f_{m_1}$, $f_{m_2}$…corresponding to $M_1$, $M_2$, …), can learn the complex, model-specific ozone sensitivities simulated by numerical models, despite their underlying uncertainties. This is challenging, as it requires $f_m$ to not only to extrapolate to $NO_x$ reduction scenarios with data cases that are not encapsulated in the training dataset, but also to be applicable to strongly divergent responses across different models (i.e., resolving $f_{m_1}$, $f_{m_2}$, … for each numerical model). Furthermore, the predictor must be available from the observations, so that the corresponding observational constraint can then be constructed. Reassuringly, surface $NO_2$ is widely observed across the measurement networks. In the second step, reconciliation, we apply the same machine learning architecture to observations to derive the observational relationship, $f_{obs}$. This relationship is then used to constrain how modeled $NO_x$-reduction-related changes in the predictor ($\Delta x_{M1}$, $\Delta x_{M2}$… i.e., $\Delta NO_2$ from model $M_1$, $M_2$, …) lead to more closely aligned ozone responses ($\Delta y_{c1}$, $\Delta y_{c2}$… the subscript $c$ here denotes constrained output).

We then conduct an additional test using the trained MOZART emulation function, $f_{MOZART}$, to predict ozone responses given the $NO_x$-emission-reduction-driven $NO_2$ changes simulated by CBMZ ($\Delta x_{CBMZ}$), and vice versa ($f_{CBMZ}(\Delta x_{MOZART})$). In simpler terms, we swap the simulated $NO_2$ changes ($\Delta x$) between the trained emulation functions of different models and use the functions with the predictor swapped to re-predict ozone responses. This predictor-swap test is designed to validate the premise that discrepancies in simulated ozone responses arise mainly from models' inconsistent chemical representations of ozone sensitivity, $f_m$, rather than differences in the simulated $NO_2$ changes in response to $NO_x$ emission reductions, $\Delta x_M$. The results for these two models, as well as the others, are available in Extended Data Fig. 1. As shown, these new predictor-swapped predictions closely mirror the original emulations. This underscores that the spread among models is driven primarily by inconsistent representations of ozone sensitivity to $NO_x$ emission changes, as approximated by $f_m$, rather than differences in the simulated $NO_2$ changes, i.e., $\Delta x_M$. Hence, this provides a foundation for employing the observational relationship, $f_{obs}$, to reconcile the modeling disagreement, explaining why models can be reconciled, as illustrated in Fig. 2.

## Using spatial neighbouring data to enhance extrapolation

The predictions in this study are generated at monthly resolution. We perform data preprocessing by averaging daily surface $NO_2$ (initially averaged between 13:00 and 14:00 local time), and daily MDA8 ozone to monthly mean values i.e., July of 2015, 2017, and 2019. The details of this conversion are documented in the observations and reanalysis products section below. This procedure is conducted for both simulated and observational data. To allow the machine learning function to extrapolate ozone responses under $NO_x$ emission reduction, we design a local spatial-expansion strategy. Specifically, when training the algorithm for each target grid cell, we increase the training sample size by including both the predictor, monthly $NO_2$, and the predictand (i.e., dependent variable), monthly MDA8 ozone, from neighbouring grid cells. Specifically, $NO_2$ and ozone from grid cells within a prescribed rectangular window (31 ×31 WRF-Chem grid cells; see example in Supplementary Fig. 2) centered on the target grid are collected to train the GPR described below. This spatial expansion broadens the range of ozone-$NO_2$ conditions available for training and enables extrapolation to $NO_x$-reduction scenarios that are not directly represented in the baseline simulations, or in the corresponding observations that are later used to construct the observational constraint for that specific target grid. For instance, a target grid cell may remain strongly $NO_x$-saturated because of high local $NO_x$ emissions and therefore may not include the transition toward a $NO_x$-limited regime. Including neighbouring grid cells that have experienced such a transition provides the additional information needed to train the emulator and observational constraint and support extrapolation.

This spatial-expansion strategy is applied iteratively to each grid across China. Compared with approaches that train a single algorithm using data from the entire study domain, our approach allows the learning function to be trained on a more localized dataset, thus retaining the locality of ozone sensitivity. Although the functions for adjacent target grid cells may share similar training samples, their different baseline $NO_2$ levels still allow distinct ozone responses to be predicted (see Supplementary Fig. 2), as demonstrated by the good emulations within the city clusters (see Fig. 1 and Extended Data Fig. 1). This strategy thus balances the need for sensitivity estimates at a local scale and the need for sufficient data to support extrapolation.

When building the observational constraint, $f_{obs}$, we do not build and predict over grid cells without sufficient neighbouring information, because extrapolation for these grids would be poorly constrained given the insufficient training data. This screening is particularly important because observations are subject to greater atmospheric variability than model output and therefore require adequate training samples. Specifically, $f_{obs}$ is not constructed for grid cells with training-sample sizes below the 25th percentile of all grid cells across China. This additional filter therefore improves the reliability of the observationally constrained estimates.

## Gaussian Process Regression

We use GPR, a Bayesian nonparametric machine learning algorithm, to model the relationship between ozone and $NO_2$ throughout this study, including model emulations, the observational constraint, and the rapid ozone sensitivity diagnosis. It is capable of smoothly fitting either linear or nonlinear relationships between ozone and $NO_2$. Instead of explicitly specifying a single deterministic functional form, GPR fits the data by finding a distribution over possible functions that are pre-defined by the prior distribution. Mathematically, a prior distribution can be expressed as below, which is then adaptively updated according to the data using Bayes' Theorem:

$$Y \sim GP(\mu, k(x, x'))$$

where $Y$ represents the vector of monthly averaged MDA8 ozone values. $\mu$ is the mean function and $k$ is the covariance function, also known as the kernel, that describes the similarity between any two given points $x$ and $x'$. Since there is only one predictor (i.e., a single dimension) used, $x$ and $x'$ thus represent two independent monthly averaged $NO_2$ concentration values. For the purpose of accentuating the $O_3$-$NO_2$ relationship, $NO_2$ is log-transformed before being converted to values relative to the baseline year (monthly averaged value in July of that year) at the target grid cell, i.e., setting log-$NO_2$ at the baseline year to 0. This conversion is important to avoid the effect of inconsistent simulated baseline $NO_2$ across different numerical models.

Following common practice in GPR, we standardize both the predictor and the dependent variable by subtracting the training-data mean and dividing by the training-data standard deviation. This sets the mean function $\mu$ to 0, allowing the GPR to be entirely explained by the kernel. Based on previously reported ozone-$NO_2$ relationships[46,54], the kernels in GPR combine a polynomial component to capture the broad ozone-$NO_2$ dependence with a squared-exponential component to represent smooth deviations from this dependence. The details of hyperparameter candidates used for cross-validation are documented in the following section.

For the model-specific emulators, i.e., $f_m$, GPR is trained using numerical-model-simulated ozone and $NO_2$ from the baseline simulations. The trained emulator then predicts ozone responses given $NO_2$ changes under the corresponding $NO_x$-reduction scenarios. For the observational constraint, $f_{obs}$, the same GPR framework is applied using observed ozone and $NO_2$, thereby allowing the observationally inferred ozone-$NO_2$ relationship to constrain the numerical-model-predicted responses.

## Training, testing, cross validation and configurations for the learning regression

The training and testing procedure in this study is designed to avoid information leakage and allow objective evaluation of the extrapolation skill. The dataset covers July of 2015, 2017, and 2019. For a target grid cell, the training samples are drawn from the neighbouring-grid dataset described above, while all data from the target grid cell itself are removed from the training set. The trained model is then applied to the withheld target grid cell to predict ozone and its response to the prescribed $NO_x$ emission-reduction scenario in that grid. For instance, when predicting ozone responses to a 30% $NO_x$ emission reduction in July 2019 for a particular target grid cell, the training data are drawn from adjacent grids within a 31 × 31 WRF-Chem grid-cell window centered on the target grid, using the available samples from July of 2015, 2017, and 2019. The trained function, either $f_m$ or $f_{obs}$, then predicts how model-simulated changes in $NO_2$ concentrations ($\Delta x_M$) given $NO_x$ reduction in that target grid translate into its ozone response. With this design, the procedure avoids spatial information leakage and further provides a stringent test of the local extrapolation capability of GPR.

Optimal hyperparameters are selected using five-fold cross-validation within the training dataset. For each fold, the training subset is standardized independently, and the resulting mean and standard deviation are then applied to the validation subset, which is the same as the standardization done in the training and testing phase. This prevents information from the validation data from entering the training procedure. The prediction error is evaluated on the validation subset using mean squared error. We pre-defined the candidates for the hyperparameters with the degrees for the polynomial kernel set to values of 1 and 2 while the variance is fixed at 1, together with initial likelihood variances of 0.01, 0.1 and 1. This setup allows the GPR model to adaptively fit the ozone-$NO_2$ relationship across different grids, considering the spatial heterogeneity of ozone sensitivity across China. For the squared-exponential component that controls the deviations from the polynomial relationship, the variance and length scale are fixed at 0.5 and 3.5, respectively. These values provided stable emulation performance across different trials while allowing smooth deviations without overwhelming the ozone-$NO_2$ dependence captured by the polynomial kernel. The parameter combination with the lowest mean validation error is selected for the final optimized model.

The optimized GPR is then refitted using the full training dataset and applied to make predictions for baseline ozone as well as ozone responses to $NO_x$ emission reductions in that specific grid and year. Given the objective of extrapolating $NO_x$-emission-driven ozone changes in this study, machine learning performance is evaluated by comparing machine-learning-predicted ozone responses to $NO_x$ reductions in the withheld grid-year test samples against those simulated by numerical models (e.g., Fig. 1). This whole procedure of training, cross-validation and testing/predicting is repeated for each grid, for either 2015 or 2019, and for each model. The same workflow is applied for the observationally constrained relationship $f_{obs}$.

## Potential discrepancies across settings in the learning regression

Like all methods, machine learning fits are not immune to uncertainty. To identify the potential discrepancies in the observational fits ($f_{obs}$) that may arise from different considerations for the setups choices of GPR setup, we conduct multiple experimental runs with different settings iteratively deviating from the aforementioned default configuration. These tests are particularly important for validating the derived observational ozone-$NO_2$ relationships, because observations are “noisy” as they are more likely to be affected by real-world factors that are not accounted for. Specifically, we iteratively preset the values for the variance in the polynomial kernel, as well as the variance and length scale in the squared-exponential kernel (see Supplementary Table 2) deviating from the default values documented above. We focus on testing and reconciling MOZART and CBMZ here, given that these are the two most divergent numerical models. Results consistent with the default constraint are produced (see Supplementary Fig. 5 and Supplementary Fig. 6), confirming the reliability of the default setups for the GPR kernels.

## Considering the meteorological effects in the observational constraint

In the main analysis, we construct $f_{obs}$ directly from the observed relationship between ozone and $NO_2$. This is motivated by the capability of the emulation function, $f_m$, which similarly derives the model-simulated ozone-$NO_2$ relationship, to replicate those strongly divergent ozone responses across models (Fig. 1 and Extended Data Fig. 1). This suggests that this relationship contains sufficient information to approximate ozone sensitivity to $NO_x$ emission changes. Nonetheless, because the real-world atmosphere is chaotic, meteorological variability may influence both ozone and $NO_2$ in observations. This may cause the observed ozone-$NO_2$ relationship, used to indicate $NO_x$-emission-driven ozone responses, to become obscured. We therefore perform an additional set of meteorology-adjusted calculations to remove the meteorological influence on the observed ozone-$NO_2$ relationship.

The meteorology-adjusted calculation adopts a residualization procedure. Under the same, monthly-averaging and spatial-expansion architecture, we fit a meteorology-only GPR to monthly mean MDA8 ozone and, in turn, to monthly $NO_2$, using three meteorological predictors: near-surface air temperature, downward shortwave radiation and relative humidity. These variables, calculated as daytime (06:00 to 18:00) daily averages and then averaged to monthly values, are selected to represent major meteorological influences on the variability of ozone and $NO_2$ while avoiding an unnecessarily high-dimensional predictor space (i.e., an overly large number of meteorological predictors). The fitted meteorological component is subtracted from ozone to obtain a residual ozone term. The same is done for $NO_2$, obtaining a residual $NO_2$ term. These residual ozone and $NO_2$ terms are then fitted with the GPR using the default configurations described above. The final prediction combines the meteorological ozone component with the ozone response inferred from the residual ozone-$NO_2$ relationship.

We repeat this procedure using a small ensemble of meteorological-kernel settings to reflect different strengths and smoothness scales of meteorological adjustment. The meteorological component for either ozone or $NO_2$ is represented with a Matérn3/2 kernel for all three meteorological variables. We iteratively run with kernel variances set at 0.05, 0.1, 0.2, 0.5, 1 and 30 while the length scale is fixed at 2. Two additional runs adopt a variance of 0.2 with length scales of 5 and 10, giving eight meteorology-adjusted runs in total. All these hyperparameters are fixed for each iterative run (i.e., set as non-trainable), specifically to allow GPR fits that deviate according to different strengths of meteorological adjustment. These perturbations thereby provide a range of supporting estimates for the default $f_{obs}$.

When meteorological effects are accounted for, similar constrained predictions are produced (as indicated by the gray density lines in Supplementary Fig. 3). Furthermore, these meteorology-adjusted constraints also show better alignment with the 2019-2015 actual observed ozone changes than all five default numerical simulations (see Supplementary Fig. 7), further supporting the reliability of the constrained predictions in the main analysis.

## Building the efficient diagnosis of ozone sensitivity

To build the efficient diagnosis, rather than simulating the full ensemble involving all five models as done to predict 30% $NO_x$-reduction-driven ozone responses, we focus on the two contrasting models, MOZART and CBMZ, with their simulations first conducted for 10%, 20%, 40% and 60% $NO_x$ reductions in July 2015 and 2019. Echoing the main text, the learning function can emulate the divergent ozone responses between the two mechanisms even under 60% $NO_x$ reductions (Supplementary Fig. 8), an important foundation for building a diagnosis across such a wide range. Additionally, given that the premise tested by swapping predictors remains valid under this extreme reduction scenario (also Supplementary Fig. 8), the resulting $NO_2$ responses from these two mechanisms can be averaged and then plugged into the observational function, i.e., $f_{obs}$ (configured with default GPR setups) to infer ozone responses across the $NO_x$ reduction range. Importantly, instead of training based on numerical model simulations[55], the rapid diagnosis here relies on the observational relationship, directly inferring ozone sensitivity of the atmosphere from observations.

Notably, the relative changes in $NO_2$ scale linearly with $NO_x$ emission reductions, and the linear fits remain similar between 2015 and 2019 (see Supplementary Fig. 9) despite differences in baseline emissions between these two years (see Supplementary Fig. 10). We therefore proceed to average the two years and interpolate $NO_2$ changes given any $NO_x$ emission reductions using the linear fit, allowing ozone responses to be estimated for other unsimulated intermediate reduction scenarios such as 15% or 55%. This further enables ozone responses to be predicted at finer reduction increments when needed. Moreover, owing to the strong linear relationship between $NO_2$ concentration changes and $NO_x$ emission reductions (see Supplementary Fig. 9), the number of current foundational simulation runs (i.e., runs with 10%, 20%, 40% and 60% $NO_x$ reductions) with the model could theoretically be reduced, saving computational costs even more substantially. Furthermore, this approach directly links $NO_x$ emission changes to ozone sensitivity predictions instead of perturbing $NO_2$ concentrations directly to derive ozone changes. Due to its localized deployment (i.e., on a grid-by-grid basis), this data-driven approach adaptively predicts ozone responses across China to the specific prescribed magnitude of $NO_x$ emission reduction in a given place, thereby moving beyond the assumption of a fixed sensitivity threshold, such as applying a predefined HCHO-to-$NO_2$ ratio to categorize ozone sensitivity regimes, which may not universally hold across space and time[56].

## Observations and reanalysis products

To train the observational constraints, $f_{obs}$, we first collect in-situ measurement data of $NO_2$ and ozone from the surface measurement network maintained by the Ministry of Ecology and Environment (MEE) in China. The design of the machine learning architecture in this study requires no further variables. This simplicity makes the constraint readily transferable and easier to deploy in other regions worldwide. Following the procedure by Weng et al.[57], we select only those stations with reliable observational records from 2015-2019 in China, giving a total of 1017 stations. To match the units of model-simulated variables, both units of ozone and $NO_2$ are first converted from ug $m^{-3}$ to ppbv, considering the change in the measurement reference state from standard conditions (273K, 1013hPa) to local ambient state (298K and 1013hPa) implemented on 1 September 2018[6]. MDA8 ozone is calculated from values of 8-hour rolling mean values at each station, with the detailed calculation procedure documented in Weng et al.[57]. For $NO_2$, we derive its daily average values based on hourly records at 13:00 and 14:00 local time, as these two hours best reflect ozone photochemistry and are also found to be closely related to MDA8 ozone[50].

With the station-based daily average data available, we spatially averaged the daily observational records within the corresponding WRF-Chem grids so that spatial resolution of the emulators, $f_m$, and the observational constraints, $f_{obs}$ are coherent. Finally, the gridded daily $NO_2$ and MDA8 ozone are averaged to monthly values. The same procedure is also applied to the simulated variables of $NO_2$ and ozone from all mechanisms in WRF-Chem.

As meteorological effects on both $NO_2$ and ozone are also considered in the additional tests for the observational constraint, we collect meteorological variables of surface temperature at 2m, surface solar radiation downward and relative humidity from ERA5 reanalysis product. These are hourly data with a 0.25° × 0.25° spatial resolution. For each meteorological variable, we average the hourly values from 06:00 to 18:00 local time for each day, converting them to daily daytime average data. We then perform a spatial linear interpolation to match values with the WRF-Chem grid. Finally, monthly averaged values are calculated from the daily values.

## Simulations with numerical models

We perform numerical simulations using WRF-Chem version 4.1.5. Five chemical mechanisms are used, MOZART, CBMZ, SAPRC, MOZART-T1 and the modified CBMZ mechanism, CB-MOD, in which the rate constants for VOC + OH reactions in CBMZ are reduced to 10% of their default values. This deliberately modified mechanism was used to validate the interpretation of the driving factors contributing to the divergent ozone responses simulated by MOZART and CBMZ in Weng et al.[30] CB-MOD produces ozone responses to $NO_x$ reductions that are closer to those from MOZART than to those from the default CBMZ mechanism (see Extended Data Fig. 1).

All simulations adopt the same WRF-Chem configuration. The model domain covers China at a horizontal resolution of 45km × 45km. The vertical grid contains 29 layers, extending from the surface layer to approximately 50 hPa. Detailed setups for the physical and dynamic components are documented in our earlier study[30]. Simulated surface ozone and $NO_2$ are obtained from the lowest model layer for training the emulators, $f_m$.

In this study, we conduct sets of simulations with numerical models for the baseline and the counterfactual $NO_x$ emission reduction scenarios. The baseline simulations aim to represent the actual atmospheric conditions of July 2015, 2017, and 2019. For each year, simulations are initialized on 28 June and run through 31 July; the first three days are discarded as spin-up, leaving July for analysis. Year-specific meteorology, chemical boundary conditions and anthropogenic emissions are used. Meteorological initial and boundary conditions are taken from the NCEP GDAS/FNL reanalysis at 1° × 1° spatial resolution and 6-hourly temporal resolution. We collect chemical initial and boundary conditions from simulations by CAM-Chem. Anthropogenic emissions within China are collected from MEIC, while emissions outside China are obtained from EDGAR.

The $NO_x$ emission reduction simulations are designed to isolate ozone responses during July 2015 and 2019 to counterfactual changes in $NO_x$ emissions. In the main perturbation experiment, $NO_x$ emissions within China are uniformly reduced by 30% relative to the corresponding baseline emissions, while meteorology and all other anthropogenic emissions are kept unchanged. All five chemical mechanisms are run for both the baseline and the 30% $NO_x$ reduction scenarios. To construct the rapid ozone sensitivity diagnosis, foundational simulations with 10%, 20%, 40% and 60% $NO_x$ emission reductions are conducted for MOZART and CBMZ.

## Data and Code Availability

The configuration files used for the WRF-Chem simulations are archived at https://github.com/adamxweng/AutoRunWRFChem415_IPR_IRR_noNudging.git (MOZART simulation as example; simulations with other chemical mechanisms share the same setups with chemical mechanism option code changed). Meteorological input data can be downloaded from the National Centers for Environmental Prediction (NCEP) Global Data Assimilation System (GDAS)/Final Analysis (FNL; 1° × 1°), accessible via https://rda.ucar.edu/datasets/ds083.2/. Simulations by Community Atmosphere Model with Chemistry (CAM-chem) are available at https://www.acom.ucar.edu/cam-chem/cam-chem. Emission data of MEIC can be obtained from: http://meicmodel.org.cn/?page_id=541&lang=en. EDGAR emission can be accessed at https://zenodo.org/record/6130621. WRF-Chem preprocessing tools mozbc, anthro_emiss, and bio_emiss can be downloaded from https://www.acom.ucar.edu/wrf-chem/download.shtml. Meteorological variables from ERA5 are available at https://cds.climate.copernicus.eu/. The template machine learning code used for the GPR-based emulation and observational-constraint analyses is archived at: https://github.com/adamxweng/MLcode_GPR_emulations_obsConstraint. All model files are available upon request.

## Extended Data Figures:

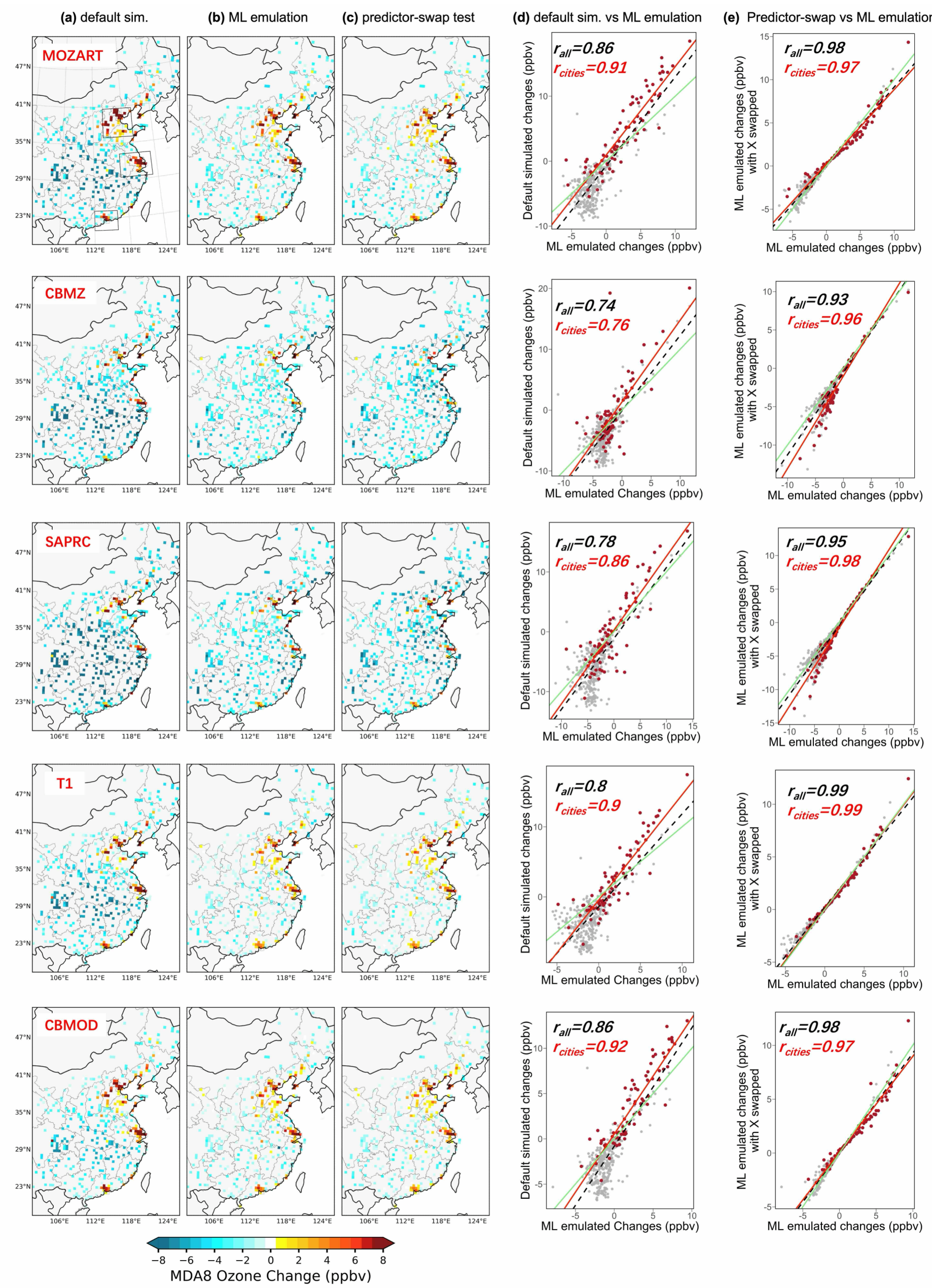

**Extended Data Fig. 1 | Emulated ozone responses to a 30% $NO_x$ emission reduction in July 2019 for all models.** Each row shows the result for a particular model with the name of model inserted in each panel in the leftmost column (column a). As in Fig. 1, the leftmost column (column a) shows the default simulations, and the second column (column b) shows the corresponding GPR emulations. The third column (column c) shows the predictor-swap tests, in which the predictor changes ($\Delta NO_2$) are exchanged between mechanisms that produces the most contrasting ozone responses (see Methods for details). The fourth column (column d) shows correlations between the default simulations (column a) and the GPR emulations (column b). Green solid lines indicate the ideal 1:1 relationship. Black dashed lines show linear fits for all scatter points, while red solid lines show linear fits for grid cells within the city clusters only. Dark-red points denote grid cells within the three city clusters, and gray points denote those outside. Pearson correlation coefficients are shown in the lower-right corner, with black values representing all grid cells across China and red values representing grid cells within the city clusters. The rightmost column (column e) shows correlations between the original emulations (column b) and the corresponding predictor-swapped test results (column c).

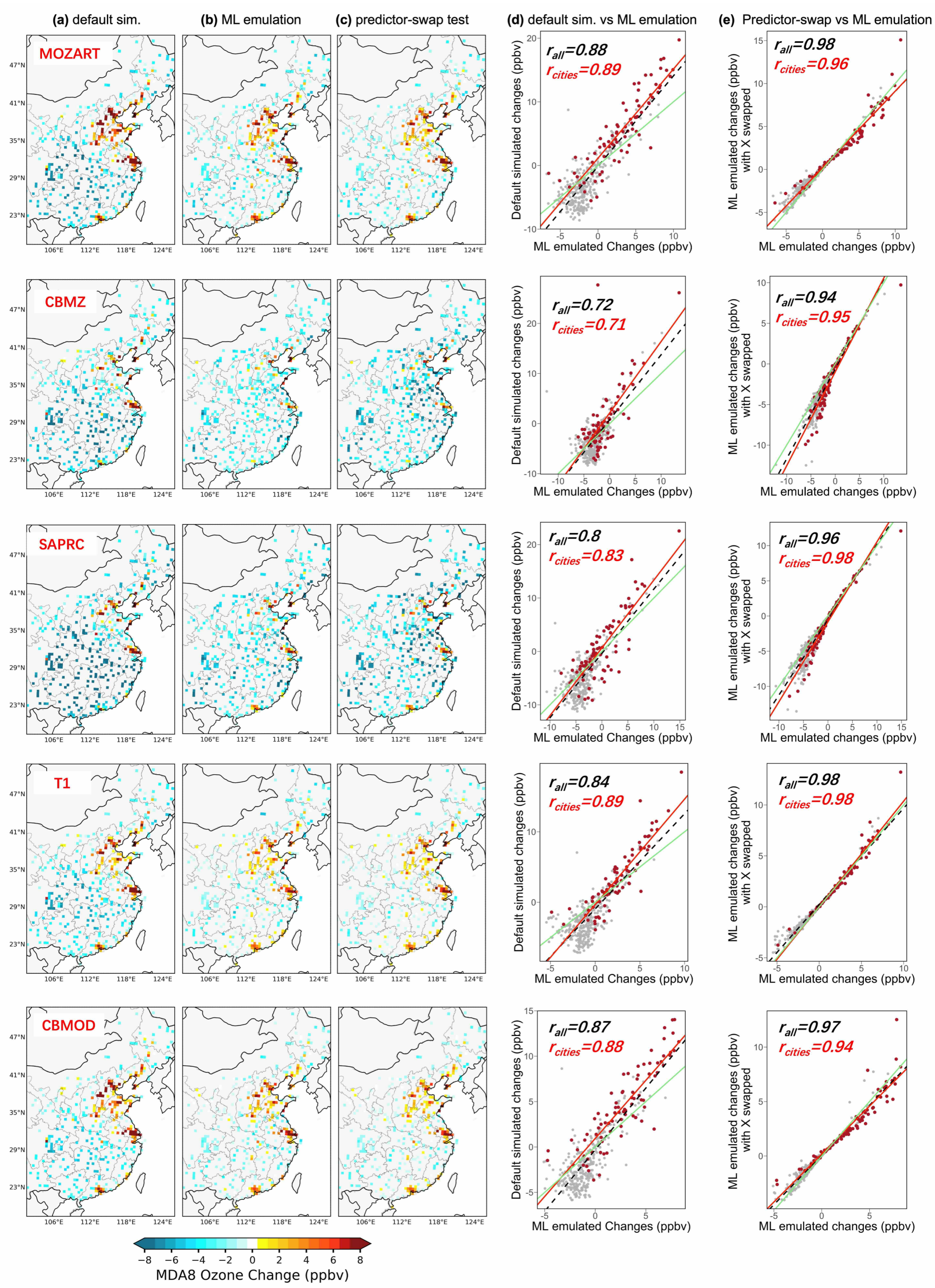

**Extended Data Fig. 2 | Same as Extended Data Fig. 1, but for a 30% $NO_x$ emission reduction in July 2015.**

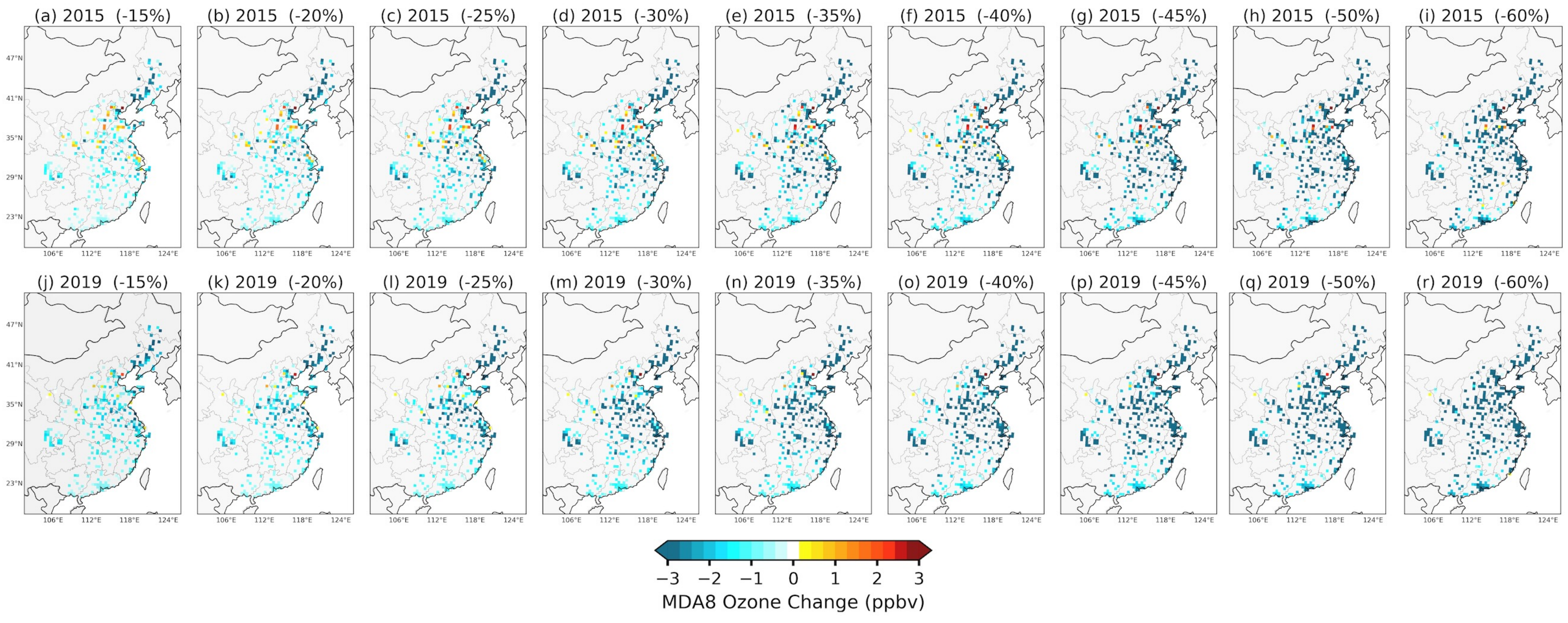

**Extended Data Fig. 3 | Predicted ozone response to counterfactual $NO_x$ emission reduction from 15% to 60%, derived using the rapid ozone-diagnosis approach, for July 2015 (top row, a-h) and 2019 (bottom row, i-p), respectively.**

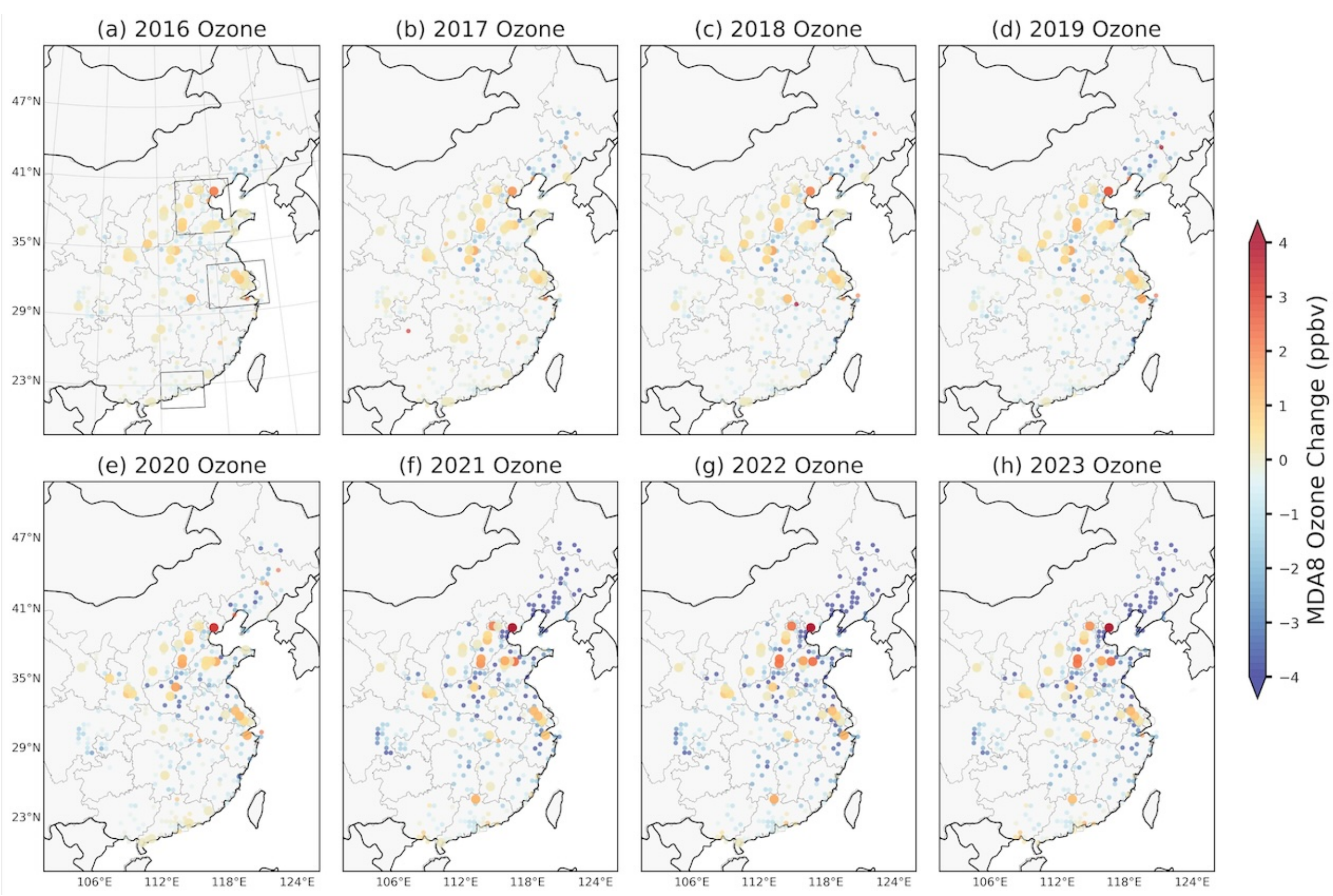

**Extended Data Fig. 4 | Predicted monthly mean MDA8 ozone changes in July 2015 when $NO_x$ emissions are changed to the levels in 2016 to 2023.** Same as Fig. 4, each point marks the centre of a WRF-Chem grid cell. Larger points indicate grid cells where ozone increases despite reductions in $NO_x$ emissions relative to July 2015.

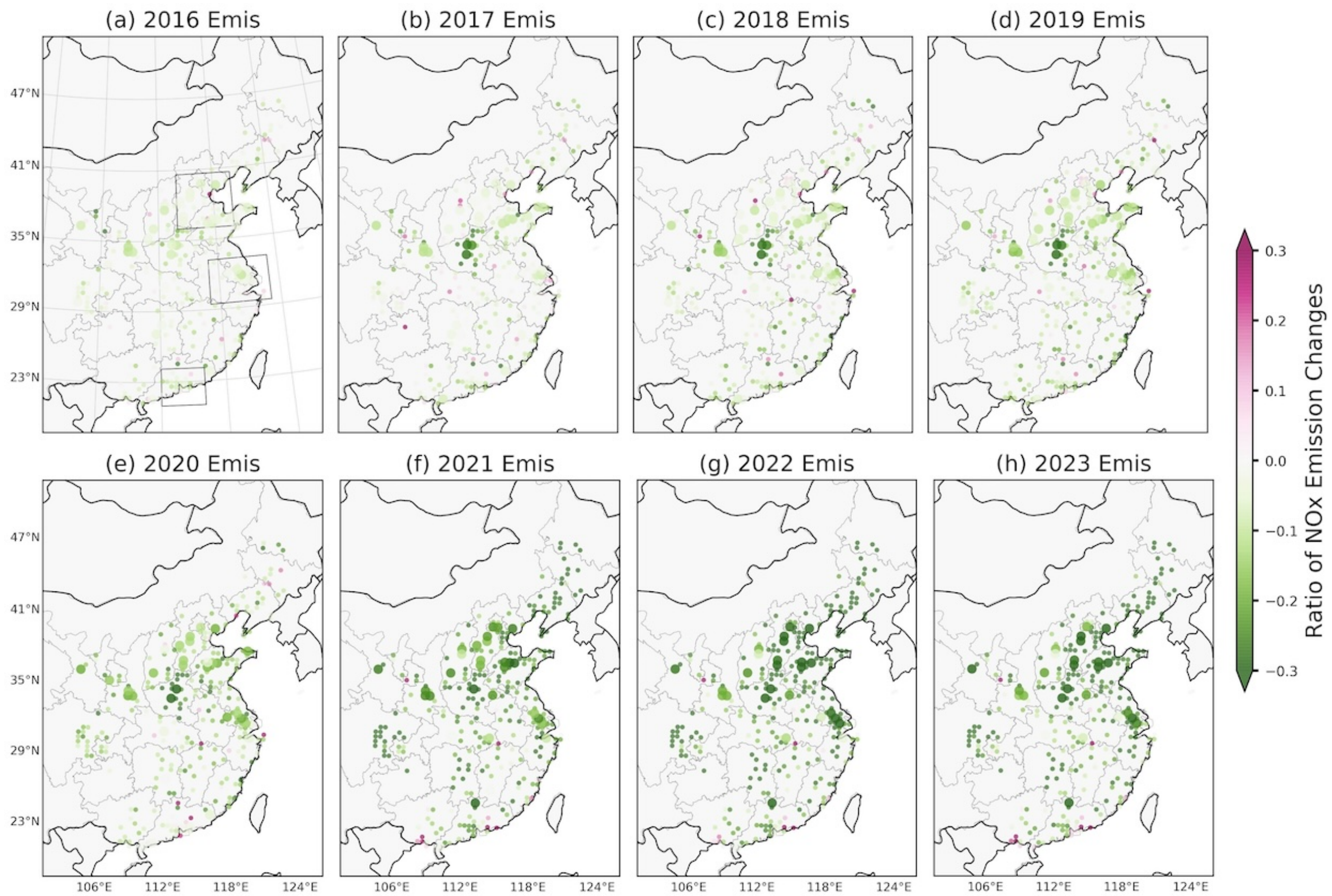

**Extended Data Fig. 5 | Relative changes in July $NO_x$ emissions from 2016 to 2023, referenced to July 2015.** Larger points denote grid cells where ozone increases despite lower $NO_x$ emissions than in July 2015, corresponding to those highlighted in Extended Data Fig. 4 and Fig. 4.

## Supplementary Figures and Tables:

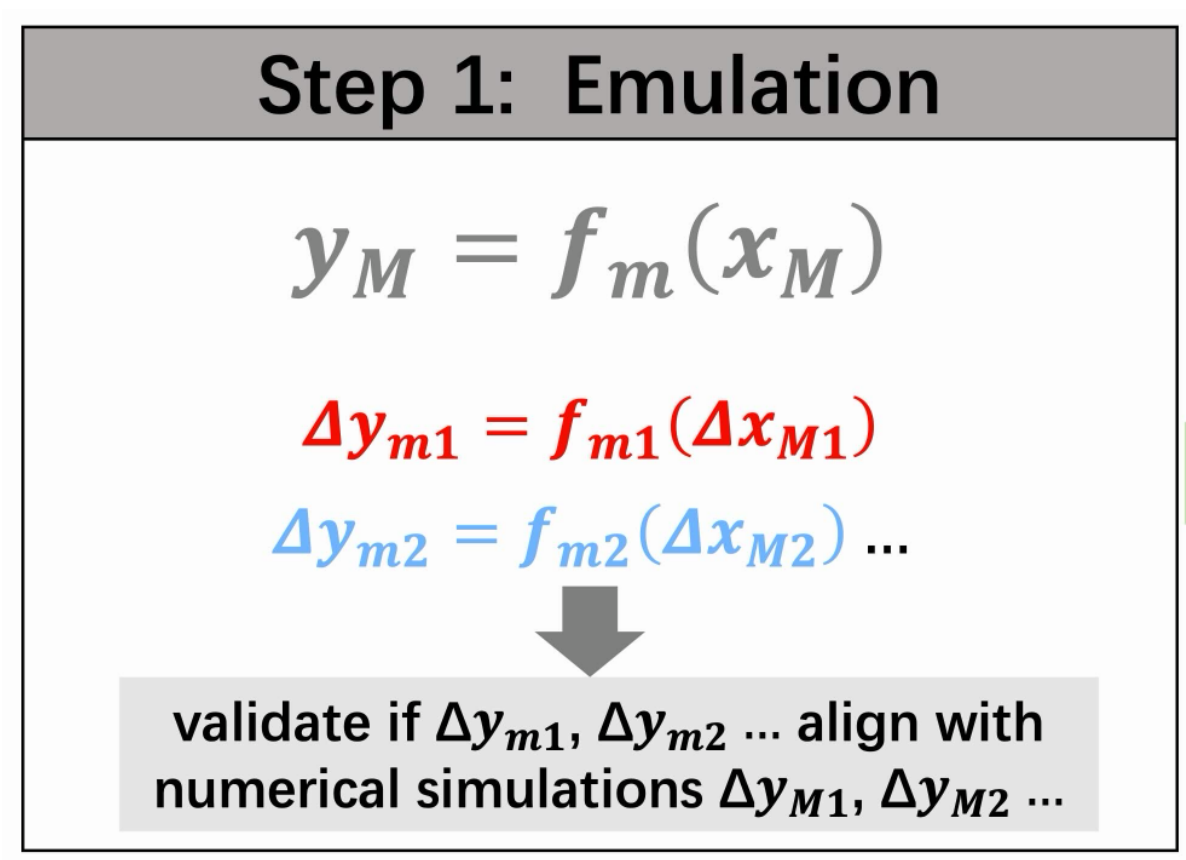

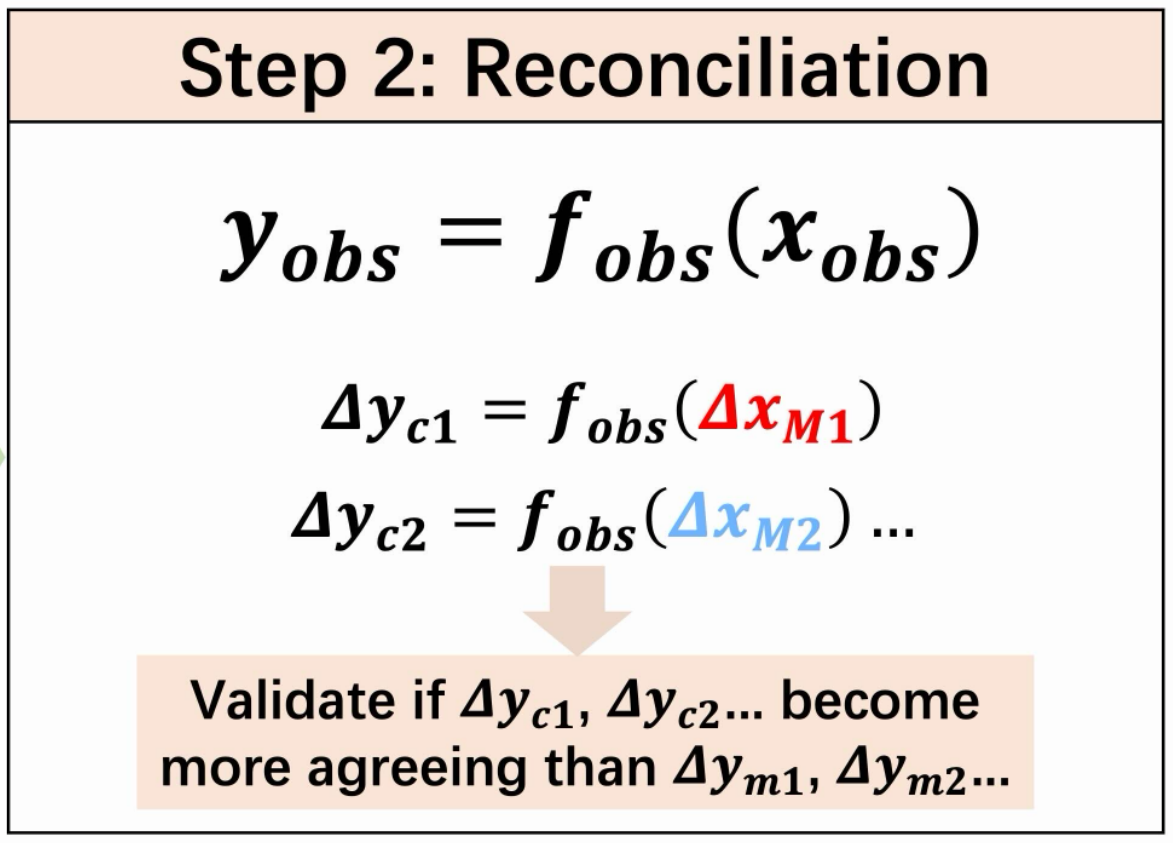

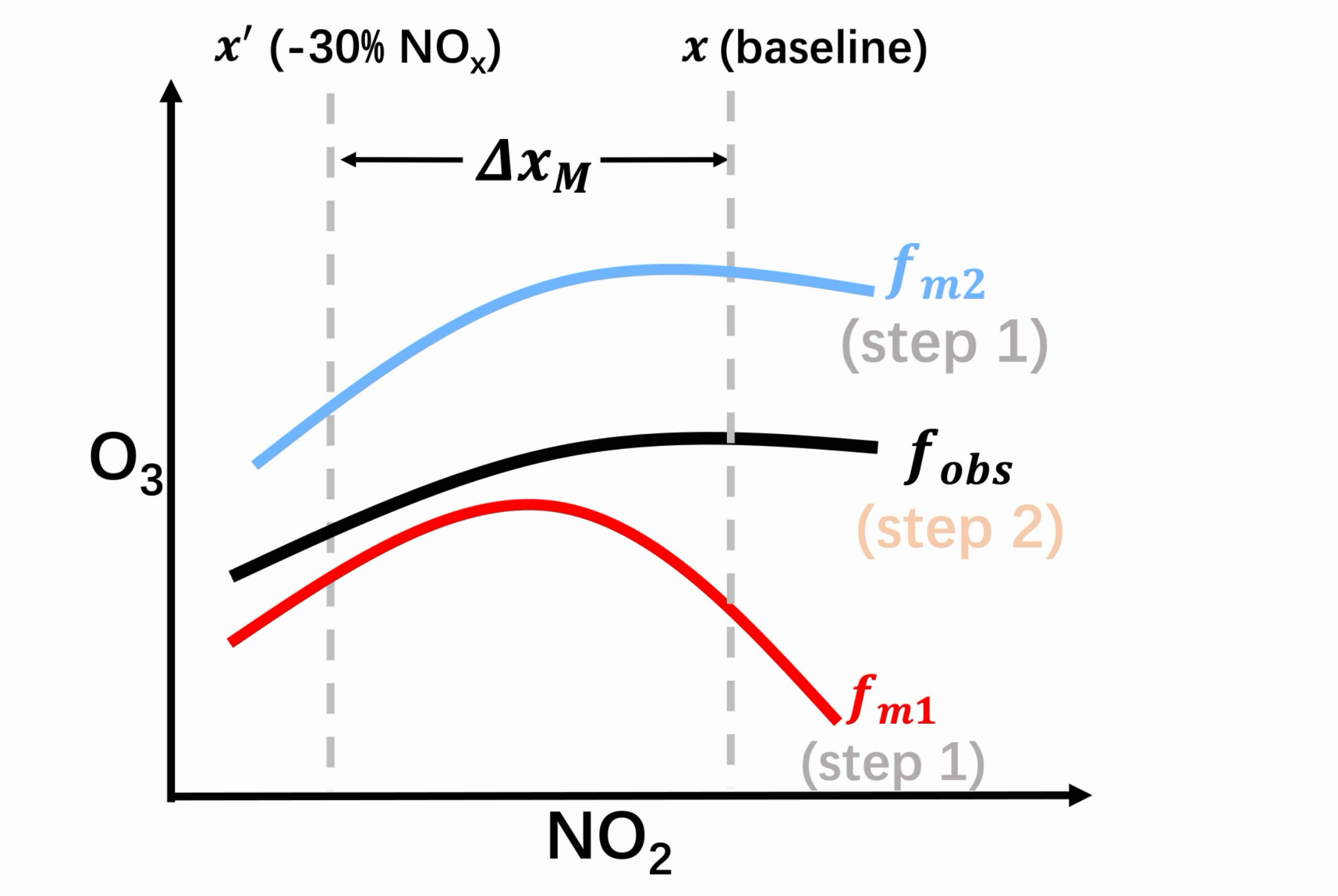

**Supplementary Fig. 1 | Overview of the machine learning framework used to emulate and constrain uncertainties in ozone sensitivity to a 30% $NO_x$ emission reduction, illustrated using two example numerical models.** $f_{m1}$, $f_{m2}$ denote the model-specific emulation functions for air-quality models 1 and 2, respectively, derived using the same machine learning architecture. $\Delta x_{M1}$, $\Delta x_{M2}$ denote model-simulated changes in the predictor, here surface $NO_2$. $f_{obs}$ represents the observational function that is derived using the same architecture as $f_m$ but trained on observational variables, i.e., observed ozone as the predictand ($y$), while observed $NO_2$ as predictor ($x$). $\Delta y_{c1}$, $\Delta y_{c2}$ denote the observationally constrained ozone responses for models 1 and 2, respectively.

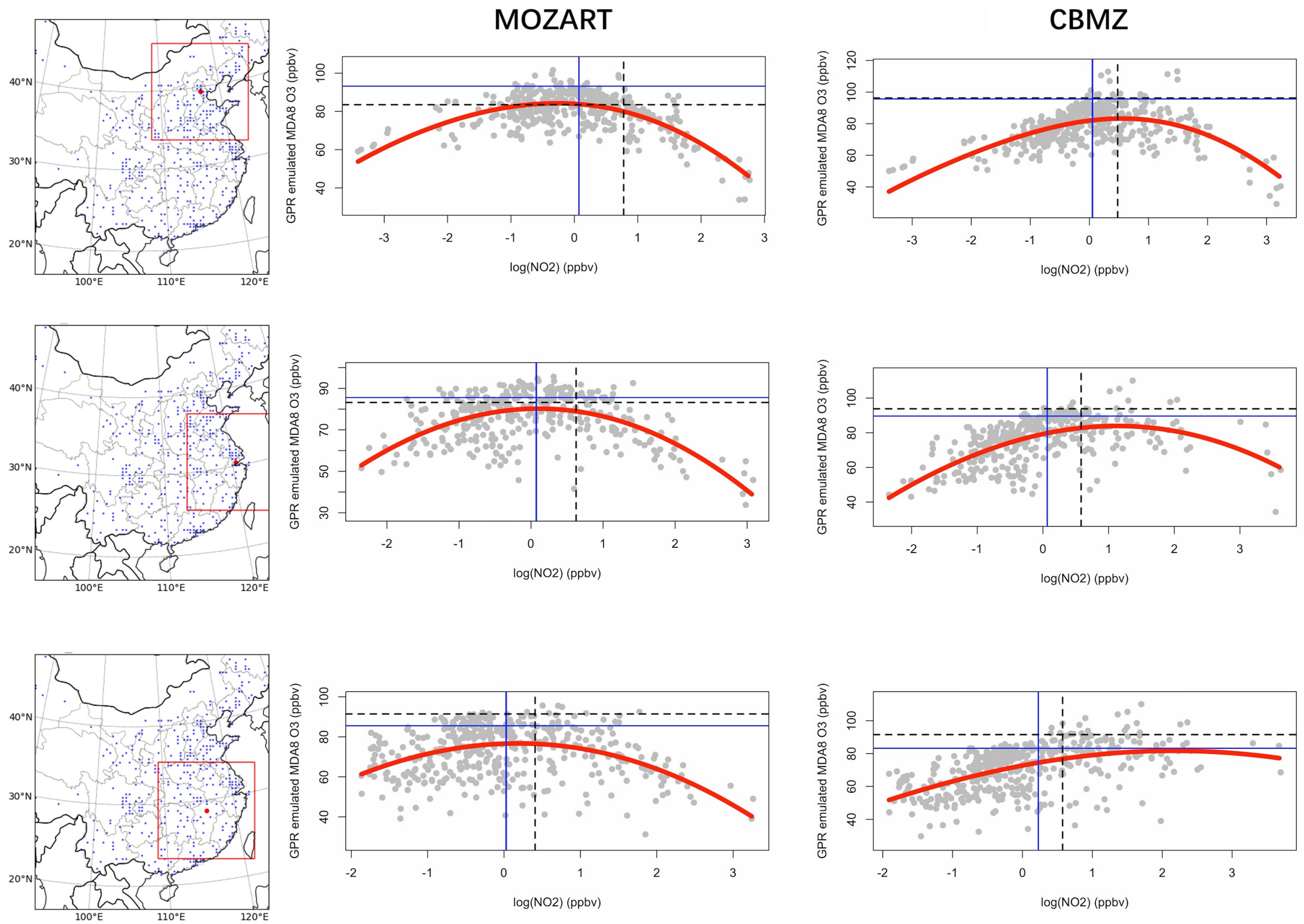

**Supplementary Fig. 2| Examples of spatial expansion strategy described in Methods and the corresponding GPR fits for emulating ozone responses to 30% $NO_x$ emission reduction in July 2019.** The left column illustrates the spatial expansion strategy. The target grid for prediction is marked with the red dot. Baseline $NO_2$ and ozone data from all the grid cells within the red rectangle, excluding the target grid, are used for training. The middle and right columns show the corresponding GPR fits for MOZART and CBMZ, respectively. Gray points in the scatter plots denote training samples from the spatial window, and red curves denote the fitted GPR relationships. Black dashed vertical and horizontal lines indicate baseline log-scaled $NO_2$ and MDA8 ozone in the target grid cell, respectively; blue dashed lines indicate the corresponding numerical-model-simulated values under the 30% $NO_x$-reduction scenario.

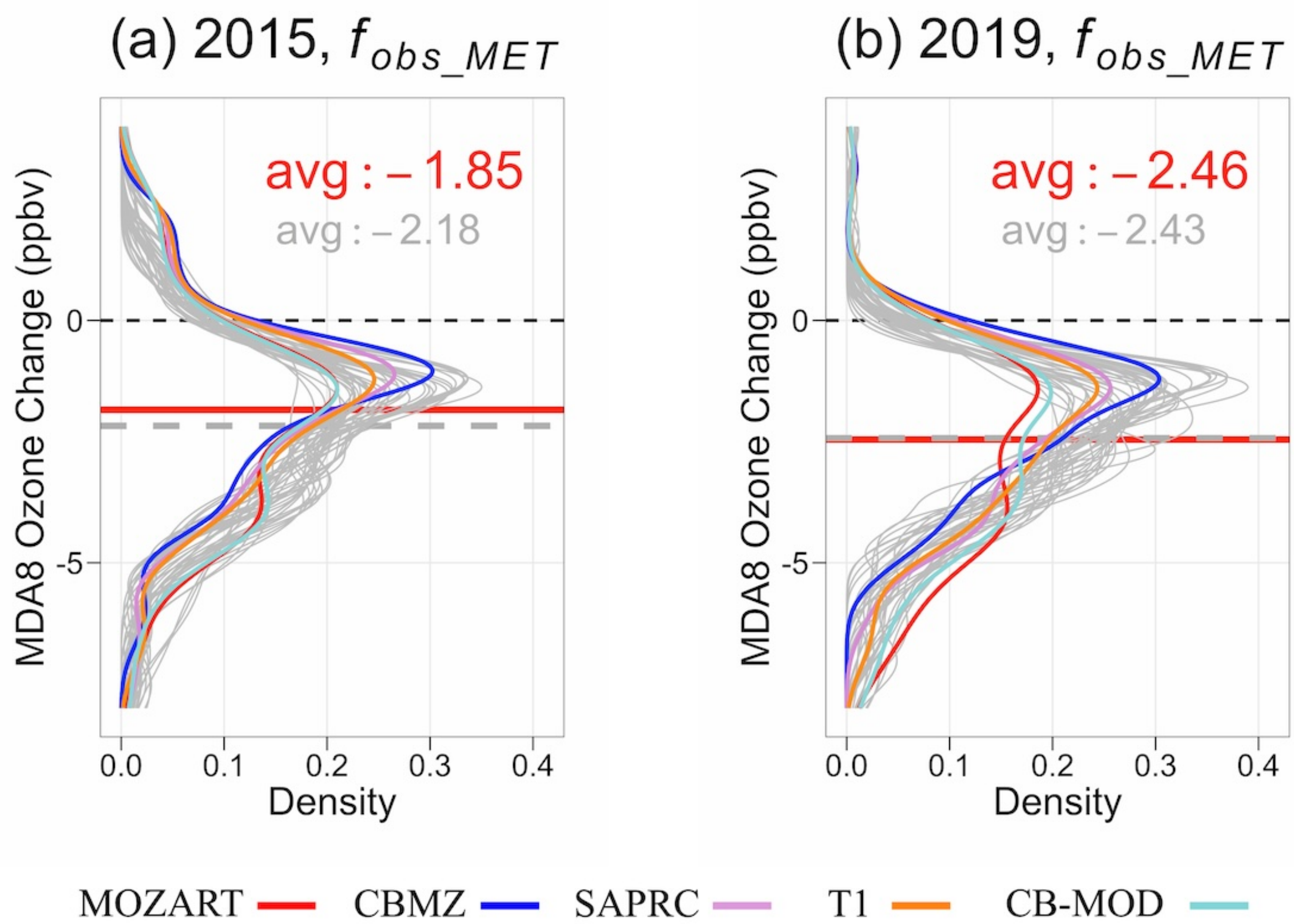

**Supplementary Fig. 3 | Density distributions for the observationally constrained ozone responses to 30% $NO_x$ emission reductions over the three city clusters with meteorological effects considered.** a,b, Observationally constrained predictions for July 2015 and 2019, respectively. The constrained results for the five numerical models using the default GPR settings are color coded, which are exactly the same as Fig. 2. The gray density distributions show constrained results by GPR with meteorological effects of temperature, solar radiation and relatively humidity included (denoted as $f_{obs_MET}$ herein; see Methods for the setups). Red horizontal lines indicate mean ozone changes by the five constrained results estimated by the default $f_{obs}$, with values shown in red at the top right corner. Whereas Gray horizontal dashed lines indicate the means of all the gray density ($f_{obs_MET}$), with average values inserted in gray at the top right corner in each panel. The black dashed horizontal lines mark the 0 value i.e., no changes in ozone given NOx reduction.

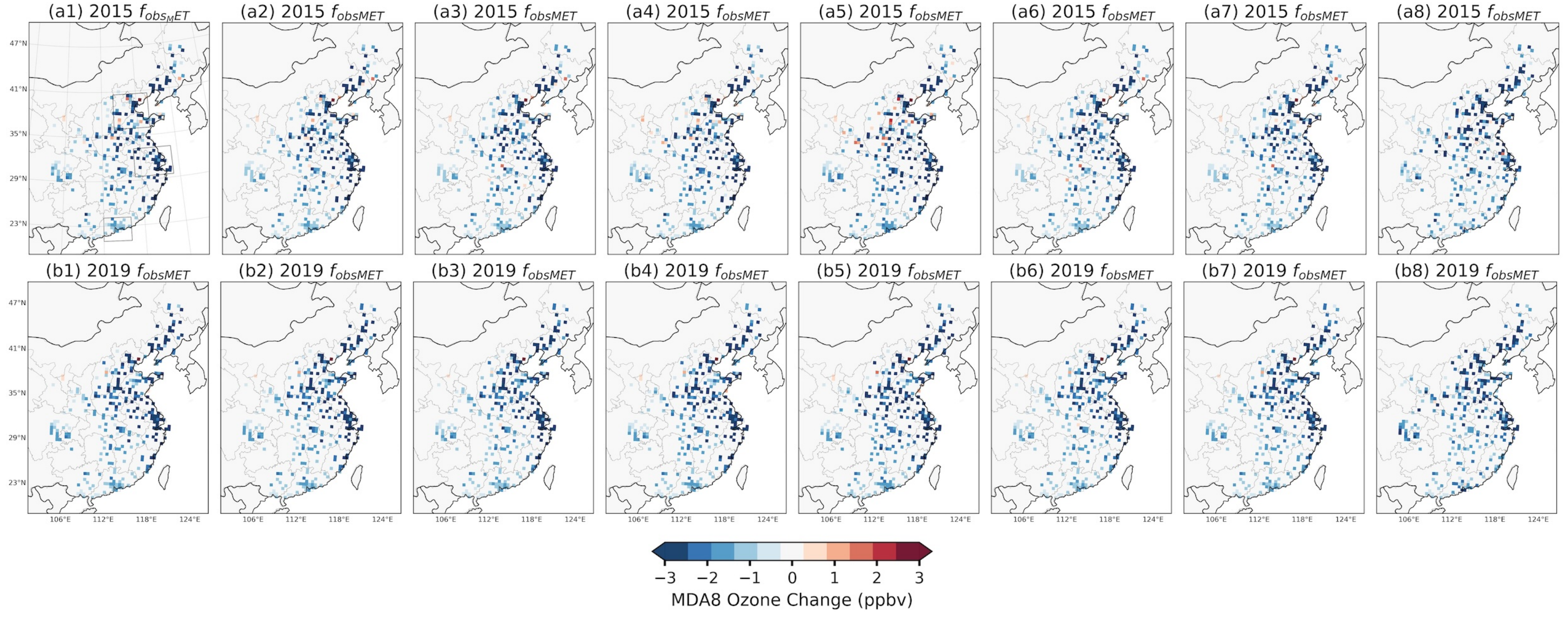

**Supplementary Fig. 4 | Ensemble-mean constrained ozone responses to a 30% $NO_x$ emission reduction in July 2015 and 2019 across the five chemical mechanisms, estimated using fobs with meteorological factors considered.** Specifically, the ozone–$NO_2$ relationship is estimated using a meteorology-adjusted framework in which the meteorologically explained components of ozone and $NO_2$ are first removed using Gaussian process regression (GPR) with three predictors (2 m air temperature, downward surface solar radiation, and relative humidity), after which the meteorological ozone component is added back to obtain the final prediction (see Methods). The top row (a1–a8) shows the July 2015 results obtained using different configurations of the Matérn 3/2 kernel to represent meteorological effects. The corresponding kernel settings are listed in Supplementary Table 1. The bottom row (b1–b8) presents the corresponding results for July 2019.

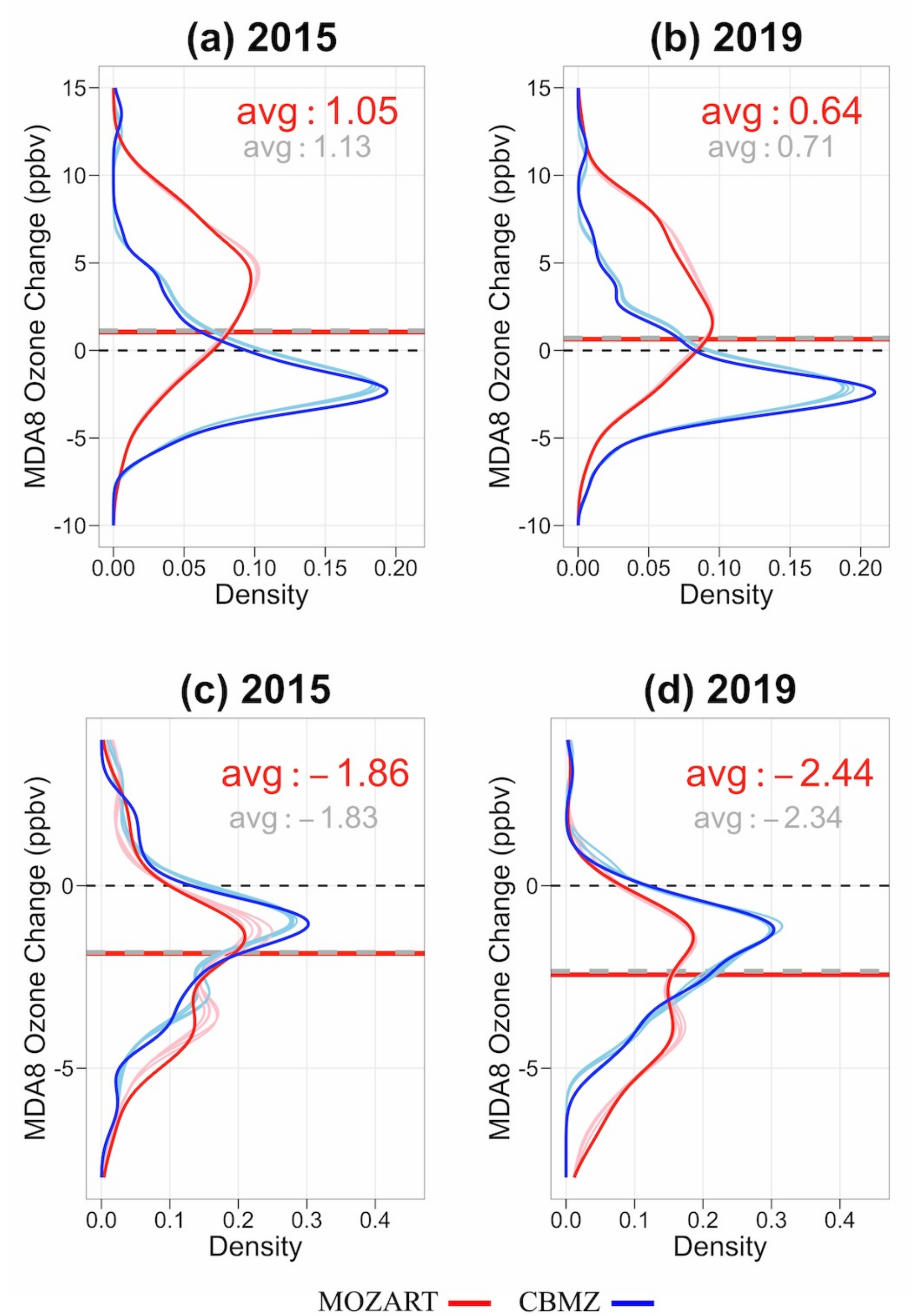

**Supplementary Fig. 5 | Density distributions for the emulated and observationally constrained ozone responses to 30% $NO_x$ emission reductions for numerical-model-simulations by MOZART (red; pink) and CBMZ (blue; skyblue) with iteratively different hyperparameter settings configured.** a,b, The emulated ozone responses within the three city clusters in July 2015 and July 2019, respectively. c,d, Corresponding observationally constrained predictions.Each pink (MOZART) or skyblue line (CBMZ) represents a different set of configuration in GPR (see Methods and Supplementary Table 2). Red horizontal lines indicate mean ozone changes by the default emulation/constraint for MOZART and CBMZ, with values shown in red at the top right corner. Gray horizontal dashed lines indicate the means of the distributions with the iterative settings, with average values inserted in gray at the top right corner in each panel.

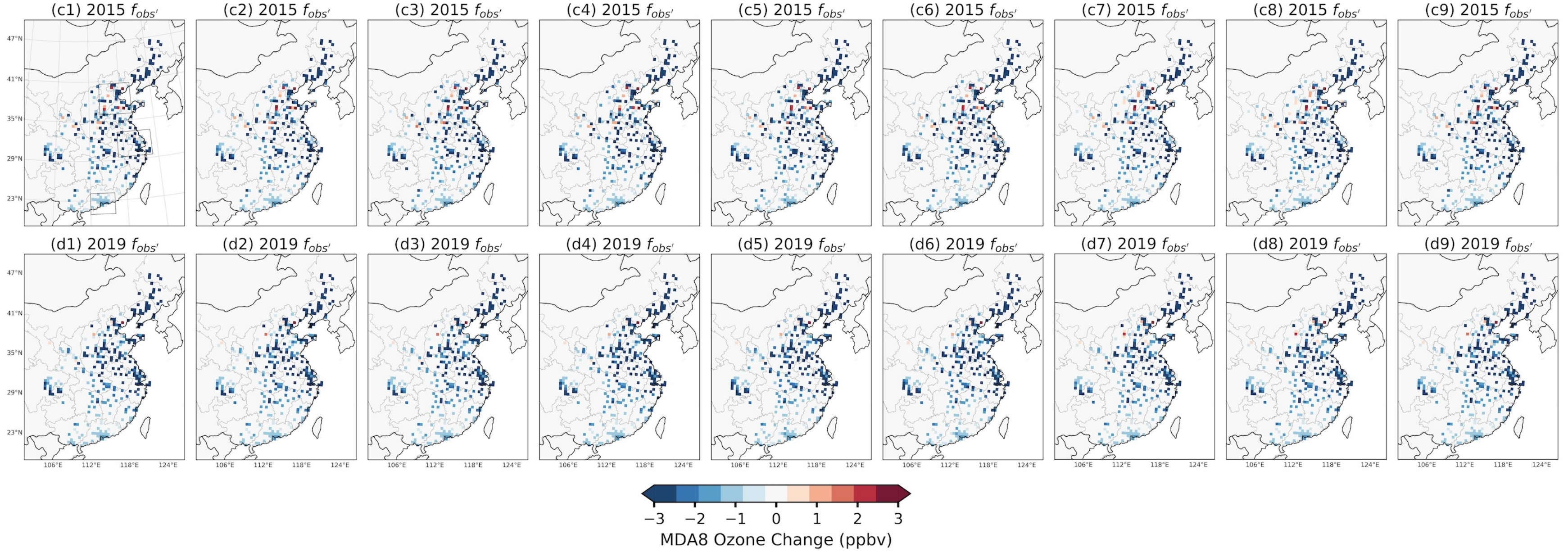

**Supplementary Fig. 6 | Mean observationally constrained ozone responses to a 30% $NO_x$ emission reduction of MOZART and CBMZ in July 2015 and 2019.** The top row (labeled with c1–a9 to avoid confusion with the runs presented in Supplementary Fig. 4) shows the July 2015 results obtained using iteratively different set of configurations for the polynomial kernel in GPR which is designed to fit the observed ozone–$NO_2$ relationship. The corresponding kernel settings are listed in Supplementary Table 2. No meteorological variables are considered here. The bottom row (d1–d9) presents the corresponding results for July 2019.

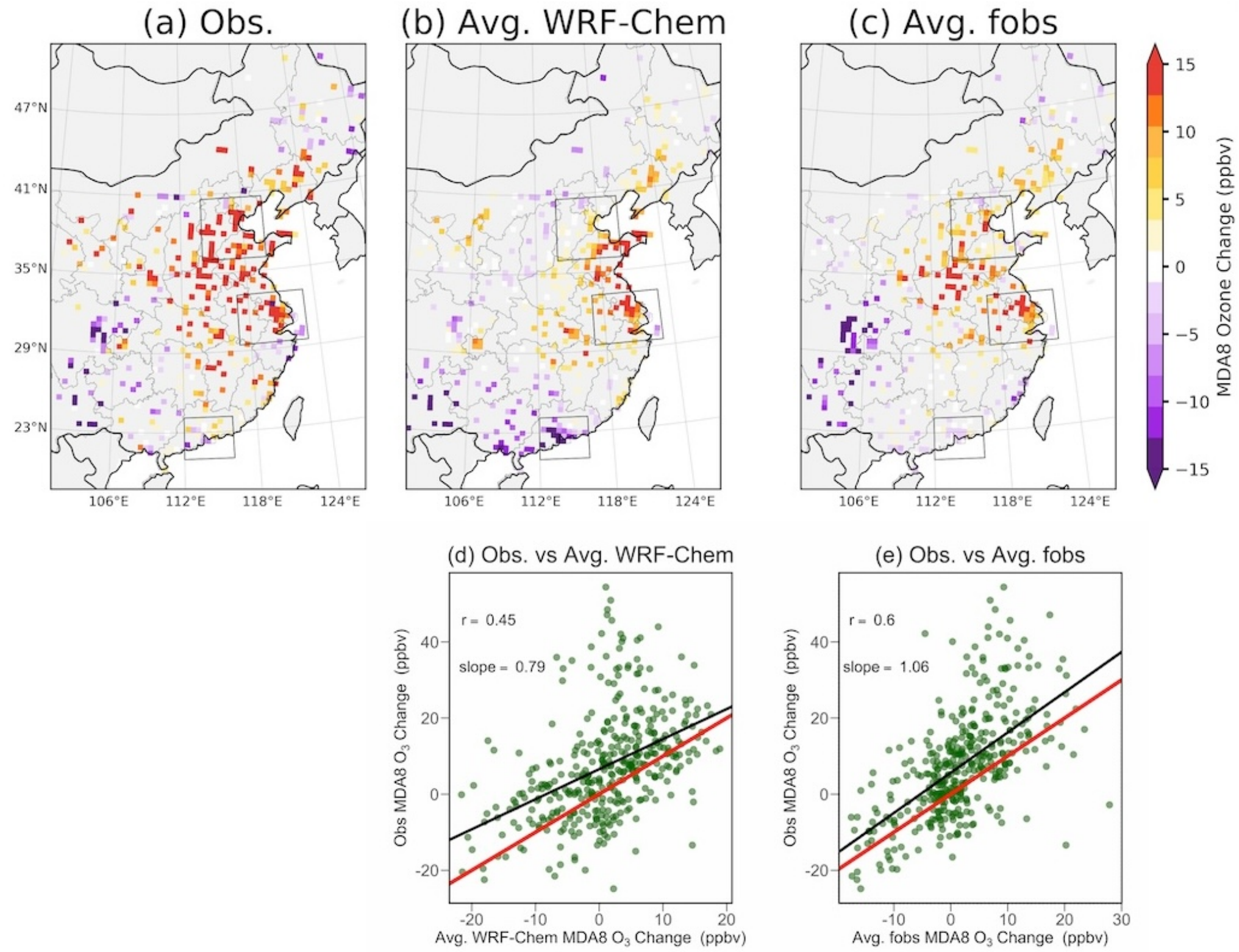

**Supplementary Fig. 7 | Comparison of the monthly mean July ozone difference between 2019 and 2015 (2019 minus 2015) from observations, the ensemble mean of the five numerical models, and the ensemble mean of $f_{obs}$ with meteorological effects from surface temperature, solar radiation and relative humidity considered (see Methods).** a, spatial distribution of the observed ozone difference. b, ensemble mean of the five models. c, ensemble mean of the observation-constrained predictions with meteorological effects considered. d, scatterplot comparing observations with the ensemble mean model results. The red line denotes the 1:1 line, and the black line shows the linear regression, with the Pearson correlation coefficient (*r*) and regression slope reported in the upper-left corner. e, same as d, but comparing observations with the ensemble mean $f_{obs}$ predictions considering the meteorological effects.

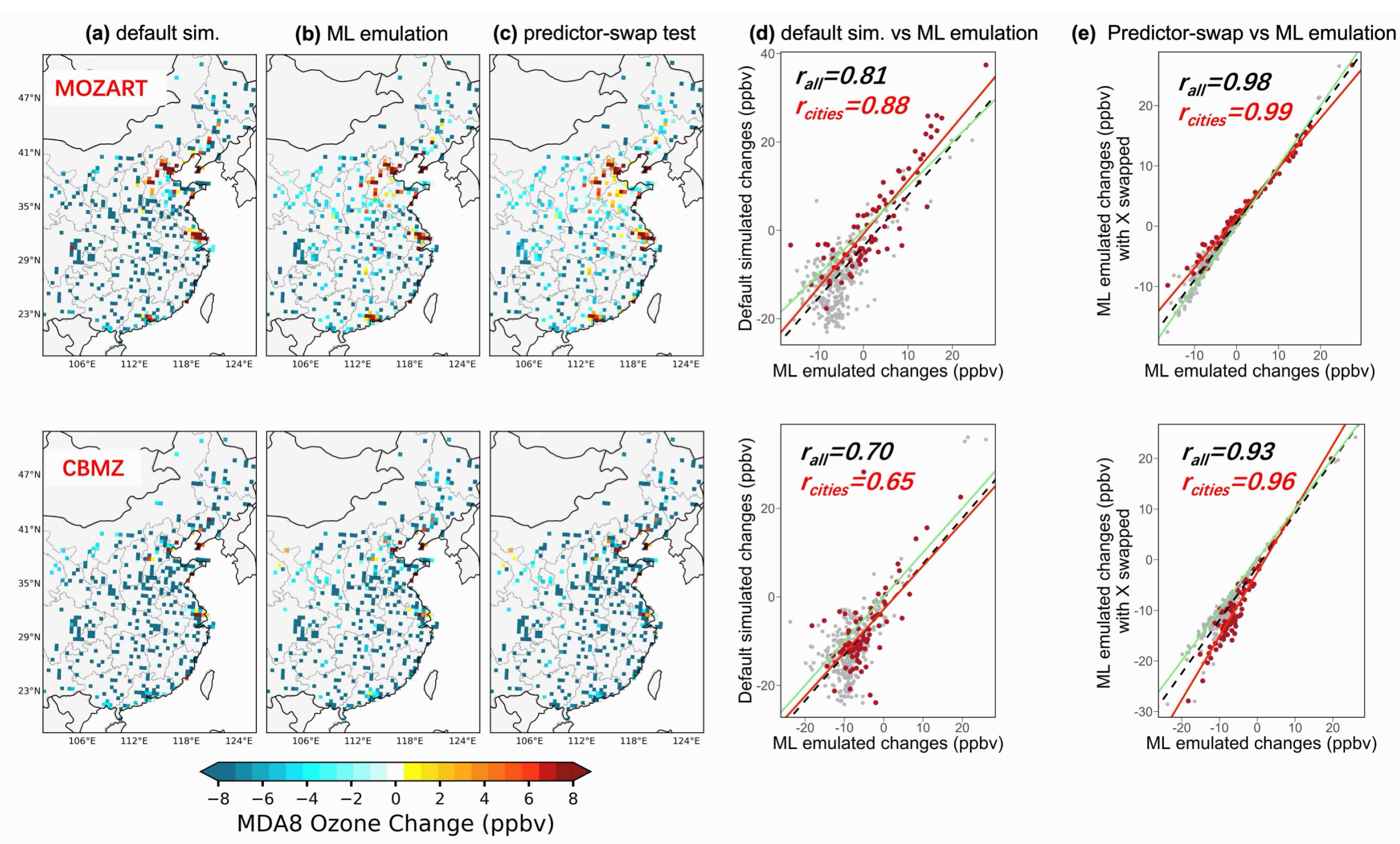

**Supplementary Fig. 8 | Emulation results for MOZART and CBMZ similar to Fig. 1 but for the scenario with $NO_x$ emission reduced by 60%.**

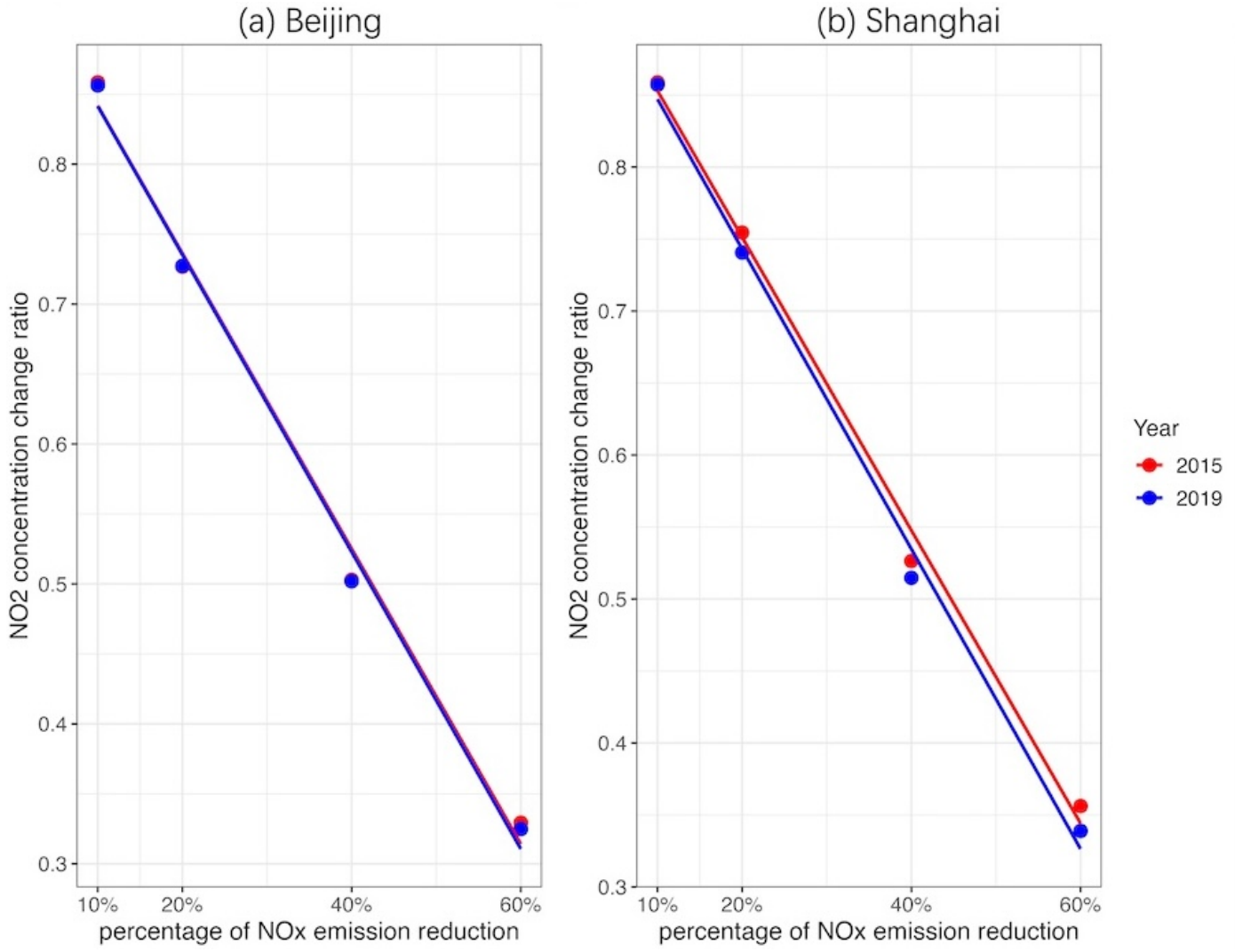

**Supplementary Fig. 9 | The relationship between NO2 concentration change ratio and percentage of $NO_x$ emission reduction in Beijing (a) and Shanghai (b) in July 2015 and 2019. Each scatter point represents the averaged value between MOZART and CBMZ.** Lines are the linear fit. Red dots and lines are July 2015 and blue are July 2019. Concentration change ratio here defined as simulated $NO_2$ concentration under a $NO_x$ reduction scenario divided by the simulated $NO_2$ concentration at the baseline.

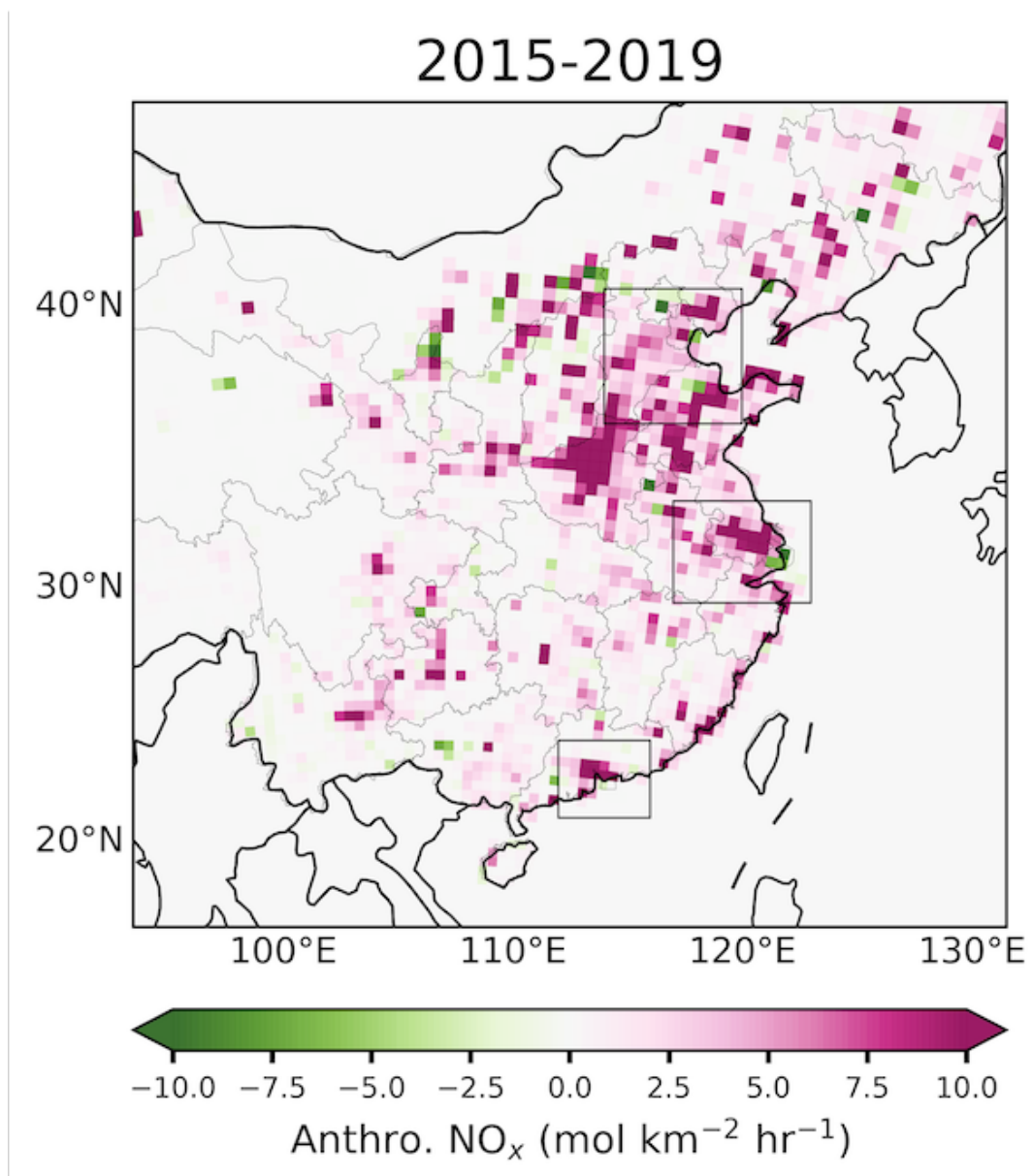

**Supplementary Fig. 10 | Difference in anthropogenic $NO_x$ emissions between July 2015 and July 2019 (2015 minus 2019).**

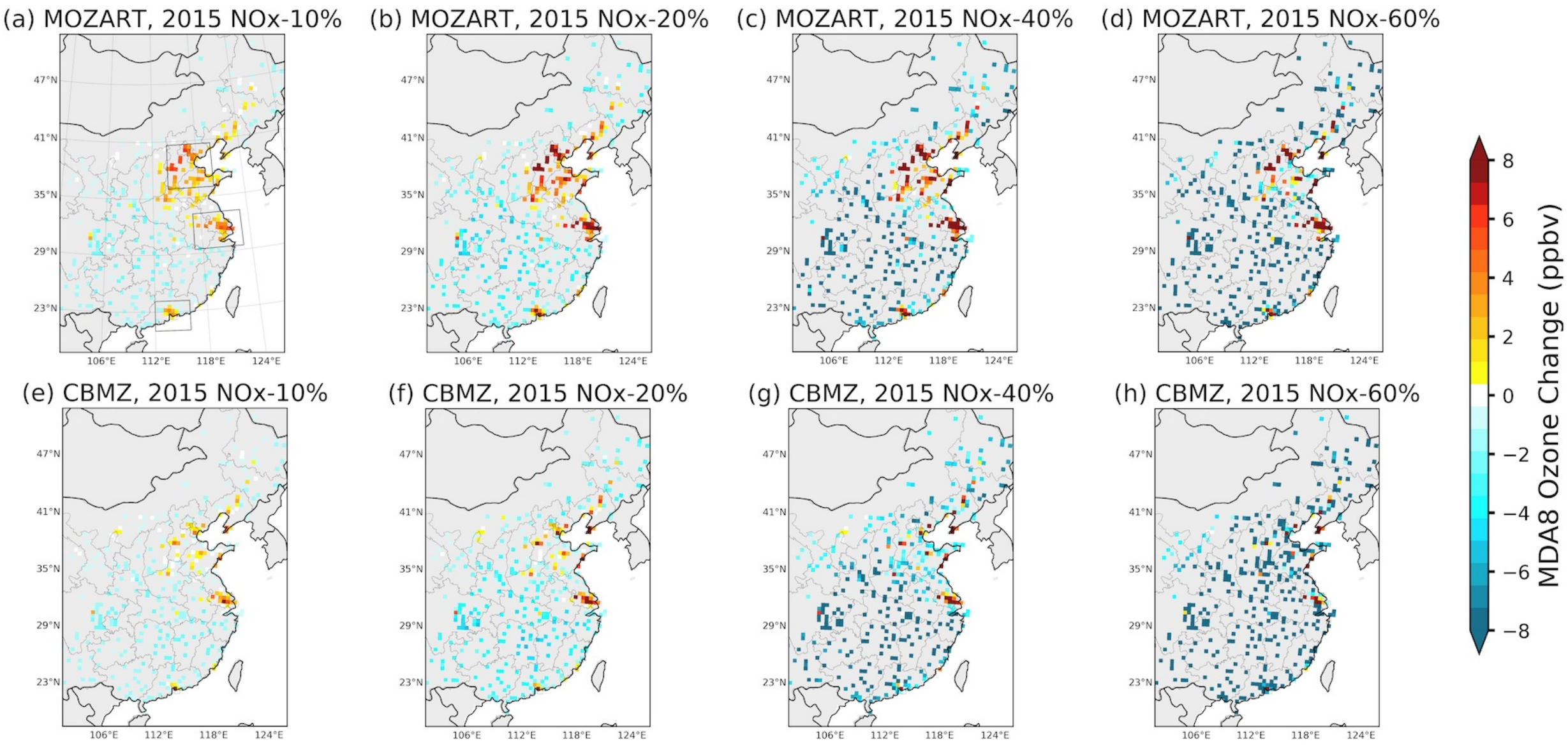

**Supplementary Fig. 11 | Default numerical-model-simulated changes in monthly mean MDA8 ozone in July under 10%, 20%, 40% and 60% $NO_x$ emission reductions.** a-d, Simulations using the MOZART mechanism. e-h, simulations by the CBMZ mechanism. The prescribed NOx reduction level is indicated in each panel title. The color scale ranges from −8 to 8 ppbv, consistent with Fig. 1, broader than that used in Fig. 3 (-3 to 3 ppbv).

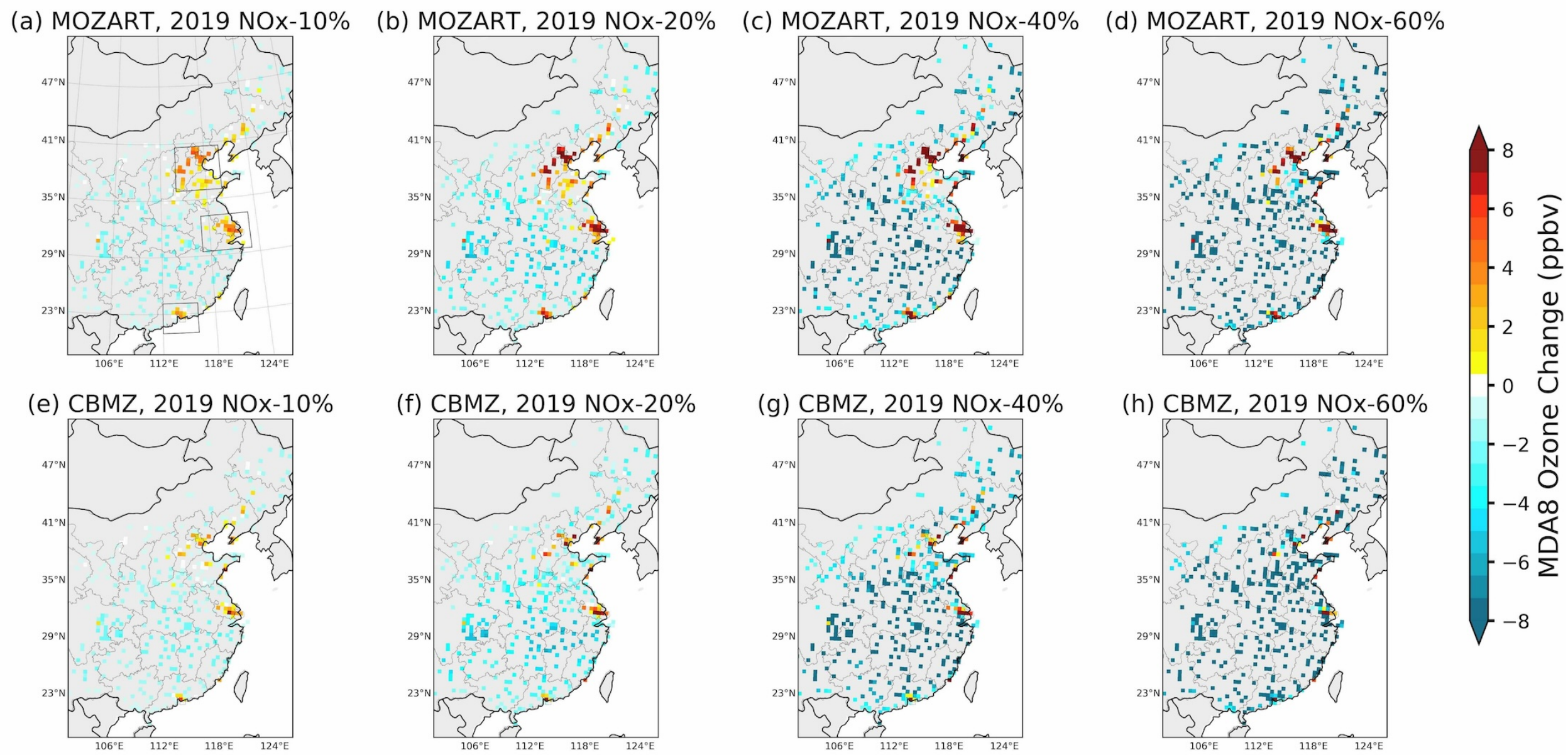

**Supplementary Fig. 12 | As in Supplementary Fig. 11, but for July 2019.**

| run ID | Variance | length scale |
|---|---|---|
| a1, b1 | 0.05 | 2 |
| a2, b2 | 0.1 | 2 |
| a3, b3 | 0.2 | 2 |
| a4, b4 | 0.2 | 5 |
| a5, b5 | 0.2 | 10 |
| a6, b6 | 0.5 | 2 |
| a7, b7 | 1 | 2 |
| a8, b8 | 30 | 2 |

**Supplementary Table 1 | Configurations of variance and length scale for Matérn3/2 kernel used to represent meteorological effects in the meteorology-only Gaussian process regression (GPR) for both ozone and $NO_2$.** The Run IDs correspond to the panel titles in Supplementary Fig. 4. For each run, both the variance and length scale are fixed (i.e., non-trainable) and are therefore not optimized during function training. This ensured that each run represented a distinct prescribed treatment of meteorological effects.

| run ID | variance for polynomial kernel | variance for squared-exponential kernel | length scale for squared-exponential kernel |
|---|---|---|---|
| c1, d1 | 5 | 0.5 | 5 |
| c2, d2 | 10 | 0.5 | 10 |
| c3, d3 | 5 | 0.5 | 10 |
| c4, d4 | 10 | 0.5 | 5 |
| c5, d5 | 10 | 0.5 | 3.5 |
| c6, d6 | 20 | 0.5 | 3.5 |
| c7, d7 | 1 | $1 \times 10^{-6}$ | 3.5 |
| c8, d8 | 1 | $1 \times 10^{-3}$ | 3.5 |
| c9, d9 | 1 | 0.05 | 3.5 |

**Supplementary Table 2 | Different sets of Configurations for variances and length scale for the polynomial and squared-exponential kernel in Gaussian process regression (GPR) for the ozone-$NO_2$ relationship.** The Run IDs correspond to the panel titles in Supplementary Fig. 6. Similar to the meteorology-adjusted run, these values are fixed (i.e., non-trainable) in each iterative run.